\documentclass [a4paper,11pt]{article}
\usepackage[english]{babel}
\usepackage{graphicx,amsmath,amsfonts,amssymb,slashed,upgreek,caption,amscd,cancel, tensor,color}
\usepackage{adjustbox}

\newcommand{\be}{\begin{equation}}
\newcommand{\ee}{\end{equation}}
\newcommand{\beq}{\begin{equation}}
\newcommand{\eeq}{\end{equation}}
\newcommand{\bea}{\begin{eqnarray}}
\newcommand{\eea}{\end{eqnarray}}

\newcommand{\DD}{\alpha}

\newcommand{\R}{R}
\newcommand{\G}{G}

\newcommand{\sR}{R}
\newcommand{\B}{{\cal B}}

\newcommand{\F}{{\cal F}}

\usepackage[stable]{footmisc}
\def\d{\delta}

\def\Mp{M_{\rm Pl}}
\def\MM{M_{*}}
\newcommand{\mfs}{{m}_4^2}
\newcommand{\tmfs}{{\tilde m}_4^2}

\def\ga{\Gamma}

\def\dkmu2{\delta K_{\mu \nu}\delta K^{\mu \nu}}
\def\pmu2{  \phi_{\mu \nu}\phi^{\mu \nu}}

\newcommand{\gammah}{h}
\newcommand{\Atwo}{{A_2}}
\newcommand{\Athree}{{A_3}}
\newcommand{\Afour}{{B_4}}
\newcommand{\Afive}{{B_5}}
\newcommand{\Bfour}{{A_4}}
\newcommand{\Bfive}{{A_5}}

\newcommand{\nbeta}{\beta}
\newcommand{\mbeta}{\beta}

\newcommand{\LS}{{{\cal A}}_K}
\newcommand{\A}{\hat{\cal A}_K}
\newcommand{\Ch}{{{\cal C}}}
\newcommand{\Co}{\hat{\cal C}}
\newcommand{\LRR}{\hat{\cal A}_R}
\newcommand{\LZ}{{{\cal A}}_R}

\newcommand{\aB}{\alpha_B}

\newcommand{\LRo}{{\cal G}}
\newcommand{\LNRo}{{\cal B}_R}

\newcommand{\C}{{\cal C}^*}
\newcommand{\LR}{{\cal G}^*}
\newcommand{\LNR}{{\cal B}_R^*}
\newcommand{\bN}{\bar N}

\begin{document}

\begin{center}
\Large{\textbf{A unifying description of dark energy}} \\[1cm] 
 
\large{J\'er\^ome Gleyzes$^{\rm a,b}$,  David Langlois$^{\rm c}$ and Filippo Vernizzi$^{\rm a}$}
\\[0.5cm]

\small{
\textit{$^{\rm a}$ CEA, IPhT, 91191 Gif-sur-Yvette c\'edex, France \\ CNRS,  URA-2306, 91191 Gif-sur-Yvette c\'edex, France}}

\vspace{.2cm}

\small{
\textit{$^{\rm b}$ Universit\'e Paris Sud, 15 rue George Cl\'emenceau, 91405,  Orsay, France}}

\vspace{.2cm}

\small{
\textit{$^{\rm c}$  APC, (CNRS-Universit\'e Paris 7), 10 rue Alice Domon et L\'eonie Duquet, 75205 Paris, France \\ 
}}
\vspace{.2cm}

\vspace{0.5cm}
\today

\end{center}

\vspace{2cm}

\begin{abstract}
We review and extend a novel approach that  we recently introduced , to describe general dark energy or scalar-tensor models. Our approach relies on an Arnowitt-Deser-Misner (ADM) formulation based on the hypersurfaces where the underlying scalar field is uniform. The advantage of this approach is that it can describe in the same language 
and in a minimal way a vast number of existing models, such as quintessence models, $F(R)$ theories,  scalar tensor theories, their Horndeski extensions and beyond. It also naturally includes Horava-Lifshitz theories. As summarized in this review, our approach provides a unified treatment of the linear cosmological perturbations about a Friedmann--Lema\^itre--Robertson--Walker (FLRW) universe, obtained by a systematic expansion of our general action up to quadratic order. This shows that the behaviour of  these linear perturbations  is generically  characterized by 
five time-dependent functions. We derive the full equations of motion in the Newtonian gauge. In the Horndeski case, we obtain the equation of state for dark energy perturbations in terms of these functions. Our unifying description thus provides the simplest and most systematic  way to confront theoretical models with current and future cosmological observations. 

\end{abstract}

\newpage 
\tableofcontents

\vspace{.5cm}

\section{Introduction}

The discovery of the present cosmological acceleration, consistently confirmed by various cosmological probes, has spurred an intense theoretical activity to account for this observational fact. Although a cosmological constant is by far the simplest explanation for this acceleration, the huge fine-tuning that 
seems required, at least from a current perspective, has motivated the exploration of  alternative models.  

As a consequence, the dark energy landscape is now very similar to that of inflation, containing  a huge number of models with various motivations and various degrees of sophistication. In fact, many of the inflationary models have been reconverted into dark energy models, and vice-versa. 
A majority of  models of dark energy, although not all of them, involve a scalar field, in an explicit or implicit way. This scalar component can be simply  added to standard gravity, like in quintessence models, or, more subtly,  intertwined with gravity itself, like in scalar-tensor gravitational theories. This illustrates the two ways to modify the dynamical equations in cosmology: either by adding a new matter component or by modifying gravity itself. 

In this paper we review and extend the  approach introduced in \cite{GLPV1} to describe in a unifying 
and minimal
way
most existing dark energy or modified gravity models that contain a single scalar degree of freedom.
This approach 
was initially inspired by the so-called effective field theory 
(EFT)  formalism, pioneered in \cite{Creminelli:2006xe,Cheung:2007st} for inflation and in \cite{Creminelli:2008wc} for minimally coupled dark energy, and later developed in the context of dark energy \cite{EFTOr,Bloomfield:2012ff,Bloomfield:2013efa} (see also \cite{PV} for a recent review and e.g.  \cite{Frusciante:2013zop,Bloomfield:2013cyf,Piazza:2013pua,Raveri:2014cka,Lombriser:2014ira}  for applications of the EFT formalism\footnote{Other general treatments of single degree of freedom dark energy, based on the equations of motion, can be found in  Refs.~\cite{Baker:2011jy,Baker:2012zs,Battye:2012eu}. 
The advantage of an action formulation is, of course, that one can easily identify ghost instabilities.
}), but exploits more systematically the 3+1 spacetime ADM decomposition by starting from a Lagrangian written only in terms of ADM quantities. This leads to an almost automatic treatment of the equations of motion, both at the background and perturbative levels. 
Our ADM approach is also at the core of several recent works \cite{Gergely:2014rna,Kase:2014baa} and is very useful for the theories beyond Horndeski that we proposed in \cite{GLPV2,GLPV3} (see also  \cite{Gao:2014soa,Kase:2014yya,Fasiello:2014aqa,Lin:2014jga,Gao:2014fra}).

In the present article, we give a slightly more general presentation of our formalism than that given in \cite{GLPV1}, by parametrizing the dynamical equations with (background-dependent) functions that are constructed directly from  partial derivatives of the initial Lagrangian with respect to the ADM tensors, rather than from partial derivatives with respect to a few scalar combinations of the ADM tensors. This makes our formalism readily applicable to a larger class of models without further preparation work, but the results are essentially the same. 
The results obtained in \cite{GLPV1} and in \cite{Bloomfield:2013efa} have been reformulated in \cite{Bellini:2014fua} by introducing dimensionless (time-dependent) functions that are combinations of those that appear in the effective formalisms previously introduced, with the advantage of clearly parametrizing deviations from General Relativity (GR).
Here we will use this notation, up to a minor redefinition and an extension to theories beyond Horndeski. 

The advantage of a unified treatment of dark energy is multiple. First, it provides  a global view of the lanscape of theoretical models, by translating them in the same language.  They thus become easy to compare, with a clear identification of approximate or exact degeneracies between the models. Moreover, a precise map also enables theorists to identify, beyond  well-known regions, unchartered  territories that remain to be explored.  A striking illustration of this is the recent realization that  theories beyond Horndeski could be free from Ostrogradski instabilities \cite{GLPV2,GLPV3}: these theories were initially  motivated by noticing that Horndeski theories correspond to a subset of all possibilities at the level of linear perturbations \cite{GLPV1}.  

Second, a unified treatment of theoretical models enormously simplifies the confrontation of these  with observational constraints. Instead of constraining separately each existing model in the literature, one can simply constrain the parametrized functions of the general formalism and then infer what this implies for each model. 
Our treatment reduces redundancies, ensuring that the number of parametrized functions is minimal for a given set of assumptions (number of space or time derivatives, etc.).
One can also identify models that are confined to ``subspaces'' of the general framework and devise optimized ways to rule them out by observations. 

Our plan is the following. 
In Sec.~\ref{sec:2}, we introduce the central  starting point of our formalism, a 
generic  Lagrangian written in the ADM formulation, and show how well-known models proposed in the literature can be reformulated in this form. In Sec.~\ref{sec:3} we rederive the main results obtained in  \cite{GLPV1}, but adopting a  more general presentation than that given originally. Then, in Sec.~\ref{sec:perts} we focus our attention on the evolution of cosmological perturbations and translate the results of the previous section into the more familiar Newtonian gauge. Moreover, we derived the perturbed Einstein and scalar field equations. In the case of Horndeski, we provide an expression for the equation of state of dark energy perturbations and discuss its observational implications. In Appendix \ref{sec:lsl} we discuss the long wavelength limit of the perturbation equations, in Appendix \ref{ref:app_syn} we give the perturbation equations in the synchronous gauge, while in Appendix \ref{sec:app_parameters} we provide the definitions of several parameters useful in the paper.

\section{A unifying action}
\label{sec:2}
\subsection{General action principle}
In this section we review the approach introduced in \cite{GLPV1}. Following \cite{EFTOr}, we assume the validity of the weak equivalence principle and thus the existence of a metric $g_{\mu \nu} $ universally coupled to all matter fields.
The fundamental  idea is then to start from a generic action that depends on the basic geometric quantities that  appear  in an ADM  decomposition of spacetime, with   {\it uniform scalar field} hypersurfaces as constant time hypersurfaces. The equations governing the background evolution and the linear perturbations can then be obtained in a generic way, up to a few simplifying assumptions (which can be easily relaxed) that are verified by most existing models.

\subsubsection{Geometrical  quantities}

Our approach relies on the existence of  a scalar field  characterized by a time-like spacetime gradient, which is a natural assumption in a cosmological context. As a consequence, the uniform  scalar field hypersurfaces correspond to space-like hypersurfaces and  can be used for a 3+1 decomposition of spacetime.

One can associate various geometrical quantities to these hypersurfaces, which will be useful in order to build a generic variational principle. 
The most immediate  geometrical quantities  are 
 the future-oriented time-like unit vector normal to the hypersurfaces $n^\mu$, which satisfies $g_{\mu\nu}n^\mu n^\nu=-1$, and 
 the projection tensor on the hypersurfaces,
 \be
 h_{\mu\nu}\equiv g_{\mu\nu}+n_\mu n_\nu\,. 
 \ee
 One can also introduce  the intrinsic curvature of the hypersurfaces, described by the Ricci tensor (which contains as much information as  the Riemann tensor for three-dimensional manifolds)
 \be
 R_{\mu\nu}\,,
 \ee
 and the extrinsic curvature tensor
 \be
 \label{extrinsic}
 K^{\mu}_{\ \nu}\equiv h^{\mu \rho} \nabla_\rho n_\nu\,.
 \ee
 Other quantities can be derived by combining the above tensors, together with the covariant derivative $\nabla_\mu$ and the spacetime metric $g_{\mu\nu}$. For example, one can define the ``acceleration'' vector field 
 \be
 a^\mu\equiv n^\lambda\nabla_\lambda n^\mu\,,
 \ee
 which is tangent to the hypersurfaces (since $n_\mu a^\mu=0$).

 With the geometrical quantities introduced above, the dependence on the scalar field is implicit. Since many dark energy models are given explicitly in terms of a scalar field $\phi$, it is useful to write down the correspondance between the various geometrical tensors and expressions of $\phi$. 
 The relation between the unit vector $n^\mu$ and the first derivative of $\phi$ is simply
 \be
 \label{n_phi}
 n_\mu= -\frac{1}{\sqrt{-X}} \nabla_\mu \phi\,, \qquad X\equiv g^{\rho\sigma}\, \nabla_\rho\phi\, \nabla_\sigma \phi \,.
 \ee
 The extrinsic curvature tensor is related to second derivatives of $\phi$, according to the expression
 \be
 \label{phimunu}
K_{\mu \nu}  =-  \frac{1}{\sqrt{-X}}\nabla_\mu\nabla_\nu\phi+n_\mu a_\nu + n_\nu a_\mu+ \frac{1}{2X} n_\mu \, n_\nu \, n ^{ \lambda} \nabla_{\lambda} X   \,,
\ee
which can be derived by substituting (\ref{n_phi}) into (\ref{extrinsic}). 

Finally, since the Lagrangian for gravitational theories often involves the four-dimensional curvature, it is useful to recall the Gauss-Codazzi relation,
\be
\label{GC1}
{}^{(4)}\! R =  K_{\mu \nu} K^{\mu \nu}- K^2 + \R  + 2 \nabla_\mu (K n^\mu - n^\rho \nabla_\rho n^\mu ) \;,
\ee
 which expresses the four-dimensional curvature ${}^{(4)}\! R$ in terms of the extrinsic curvature tensor and of the 
 intrinsic curvature. We will always denote the four-dimensional curvature with the superscript $(4)$ to distinguish it from the hypersurface intrinsic curvature.

 \subsubsection{ADM coordinates}
So far, all geometrical quantities have been introduced intrinsically, without reference to any specific coordinate system. However, since spacetime is endowed with  a preferred slicing, defined by the uniform scalar field hypersurfaces,  it is convenient to use coordinate systems especially adapted to this slicing, in other words so that constant time hypersurfaces coincide with the preferred hypersurfaces.

We thus express the four-dimensional metric  in the ADM form  
  \be
\label{ADM}
ds^2=-N^2 dt^2 +{h}_{ij}\left(dx^i + N^i dt\right)\left(dx^j + N^j dt\right) \, ,
\ee
where $N$ is the lapse and $N^i$ the shift.  In matricial form, the components of the metric and of its inverse are given respectively by 
\beq
g_{\mu\nu}=
\left(
\begin{array}{cc}
-N^2+ h_{ij}N^i N^j & h_{ij}N^j
\\
h_{ij}N^i & h_{ij}
\end{array}
\right)\,,
\quad
g^{\mu\nu}=
\left(
\begin{array}{cc}
-{1}/{N^2}& {N^j}/{N^2}
\\
{N^i}/{N^2} & h^{ij}-{N^i N^j }/{N^2}
\end{array}
\right)\,.
\eeq
In ADM coordinates, we obtain 
\be
X=g^{00}\dot\phi^2(t)=-\frac{\dot\phi^2(t)}{N^2}\,,
\ee
since the scalar field depends only on time, by construction.
The components of the normal vector are thus given by 
\be
n_0=-N \,, \qquad n_i=0\,.
\ee
 The components of the extrinsic curvature tensor can be written as
 \be
\label{extrinsic_ADM}
K_{ij} = \frac{1}{2N} \big(\dot h_{ij} - D_i N_j - D_j N_i \big) \;,
\ee
where a dot stands for a time derivative with respect to $t$, and $D_i$ denotes the covariant derivative associated with the three-dimensional spatial metric $h_{ij}$. Spatial indices are lowered and raised by the spatial metric.

 In the following, we will consider general gravitational actions which can be written in terms of the geometrical quantities that we have introduced, expressed in ADM coordinates, 
 \beq
\label{start_point}
S_g=\int d^4 x \sqrt{-g}\,  L(N,  K_{ij}, R_{ij}, h_{ij},D_i ;t) \;, 
\eeq
with $\sqrt{-g}=N\sqrt{h}$, where $h$ is the determinant of $h_{ij}$. Note that, by construction, the above action is automatically invariant under spatial diffeomorphisms, corresponding to a change of spatial coordinates. 
 
 \subsection{Examples}
 To make things concrete, let us illustrate our formalism by listing briefly the main  scalar tensor theories that have been studied in the context of dark energy  and by presenting their explicit reformulations  in the general form (\ref{start_point}).

 \subsubsection{General relativity }
Before introducing models with a scalar component, let us start by simply rewriting the action for  general relativity in the above ADM form. 
Starting from the Einstein-Hilbert action
\be
S_{\rm GR}=\int d^4x \sqrt{-g}\, \frac{M_{\text{Pl}}^2 }2\, {}^{(4)}\!R\,,
\ee
and substituting the Gauss-Codazzi expression (\ref{GC1}), one can get rid of the total derivative term and express the action in terms of the extrinsic and intrinsic curvature terms only.  Therefore, one easily obtains a Lagrangian of the form (\ref{start_point}) for General Relativity (GR), which reads
  \be
 \label{pureGR}
 L_{\text{GR}}=\frac{M_{\text{Pl}}^2}2\left[ K_{ij}K^{ij}-K^2+\R \right]\,.
 \ee
 Note that, in contrast with the following examples that intrinsically contain a scalar degree of freedom, the slicing of spacetime is arbitrary since there is no preferred family of spacelike hypersurfaces. This means that the Lagrangian (\ref{pureGR}) contains an additional symmetry, leading to full four-dimensional invariance, which is not directly manifest in the ADM form.

 \subsubsection{Quintessence and $k$-essence}
 The simplest way to extend  gravity with a scalar component is  to add   to the Einstein-Hilbert action a standard action  for the  scalar field, which consists of a kinetic term plus a potential. This corresponds to 
 quintessence models.   The initial covariant action
 \be
 S=S_{\rm GR}+ \int d^4x \sqrt{-g}\left(-\frac12\partial_\mu \phi\partial^\mu \phi -V(\phi)\right)
 \ee
 leads to the ADM Lagrangian
 \be
 \label{Quintessence}
 L=L_{\rm GR}+L_{\rm Q}\,, \qquad 
L_{\text{Q}}(t,N)=\frac{\dot\phi^2(t)}{2N^2} -V(\phi(t))\,.
 \ee
 In a similar way, one can describe $k$-essence theories~\cite{ArmendarizPicon:2000dh,ArmendarizPicon:2000ah}
   by expressing their Lagrangian $P(X,\phi)$ in terms of $N$ and $t$:
 \be
 L_{ k\text{-essence}}(t,N)=P\Big[-\frac{\dot\phi^2(t)}{2N^2}, \phi(t)\Big]\,.
 \ee

 \subsubsection{$F({}^{(4)}\!R)$ theories}
 Theories described by a Lagrangian consisting of a nonlinear function of the four-dimensional  curvature scalar ${}^{(4)}\!R$ are equivalent to a scalar-tensor theory.  Indeed, it is easy to verify that the Lagrangian 
  \be
 \label{f(R)}
L_{F(R)}= F(\phi)+F_\phi(\phi)({}^{(4)}\!R-\phi)\, ,
 \ee
 is equivalent to the Lagrangian $F({}^{(4)}\!R)$, as they lead to the same  equations of motion (as long as $F_{{}^{(4)}\!R {}^{(4)}\!R}\neq 0$). Given this property, one can then use eq.~\eqref{GC1} to rewrite the above Lagrangian, after integration by parts, in  the ADM  form
  \be
 \label{f(R)ADM}
L_{F(R)}=F_\phi(\R + K_{\mu \nu} K^{\mu \nu}- K^2)+2F_{\phi\phi}\, K\sqrt{-X} + F(\phi)- \phi F_\phi\, .
 \ee

 \subsubsection{Horndeski theories}
 In the last few years, a lot of activity has been focussed on a large class of theories, known as Hordenski theories \cite{horndeski},  
 shown to be equivalent to 
 Generalized Galileons \cite{Deffayet:2011gz} in   \cite{Kobayashi:2011nu}. 
Although their Lagrangians contain up to second derivatives of a scalar field, these theories correspond to  the most general scalar-tensor theories that directly lead to at most second order equations of motion. As such, they include all the examples introduced above. They can be written as an arbitrary linear combination of the following Lagrangians:
 \begin{align}
L_2^{H} [G_2] \equiv & \; G_2(\phi,X)\;,  \label{L2} \\
L_3^{H} [G_3] \equiv & \; G_3(\phi, X) \, \Box \phi \;, \label{L3} \\
L_4^{H} [G_4] \equiv &\;G_4(\phi,X) \, {}^{(4)}\!R - 2 G_{4X}(\phi,X) \big[ (\Box \phi)^2 - ( \nabla^\mu \nabla^\nu \phi )(\nabla_\mu \nabla_\nu  \phi ) \big] \;, \label{L4} \\
L_5^{H} [G_5] \equiv & \;G_5(\phi,X) \, {}^{(4)}\!G_{\mu \nu} \nabla^\mu \nabla^\nu \phi +\frac13 G_{5X} (\phi,X)  \times \nonumber \\
& \big[ (\Box \phi)^3 - 3 \, \Box \phi \, (\nabla^\mu \nabla^\nu \phi )(\nabla_\mu \nabla_\nu  \phi )+ 2 \, 
(\nabla_\mu \nabla_\nu \phi  )(
\nabla^\sigma \nabla^\nu \phi )(
\nabla_\sigma \nabla^\mu \phi) \big] \;.\label{L5}
\end{align}
Rewriting these Lagrangians in the ADM form turns out  to be significantly more involved 
than in
the previous examples. This calculation was undertaken in \cite{GLPV1}, where all the details are given explicitly. The final result is that the above Lagrangians \eqref{L2}--\eqref{L5} yield, in the ADM form, combinations of the following four Lagrangians
 \begin{align}
L_2^{H} = & \; F_2(\phi,X)\;,  \label{L2ADM} \\
L_3^{H}= & \; F_3(\phi, X) \, K\;, \label{L3ADM} \\
L_4^{H}  = &\;F_4(\phi,X) \, \R +(2 XF_{4X}-F_4) (K^2 - K^{ \mu \nu} K_{ \mu \nu}) \;, \label{L4ADM} \\
L_5^{H}= & \;F_5(\phi,X) \, \G_{\mu \nu} K^{\mu \nu}  -\frac13  X F_{5X}   (K^3 - 3 K K_{\mu \nu}K^{\mu \nu} + 2  K_{\mu \nu}  K^{\mu \sigma} K^\nu_{\ \sigma}) \;. \label{L5ADM}
\end{align}
The functions $F_a$ appearing here are related to the $G_a$ in eqs.~\eqref{L2}--\eqref{L5} through (see \cite{GLPV1} for details)
\be
\begin{split}
F_2=& G_2-\sqrt{-X}\int \frac{G_{3\phi}}{2\sqrt{-X}}\, dX\, , \\
F_3=&-\int G_{3X}\sqrt{-X}\,dX-2\sqrt{-X}G_{4\phi}\, ,\\
F_4=& G_4+\sqrt{-X} \int\frac{ G_{5\phi}}{4\sqrt{-X}}\, dX\, ,\\
F_5=&-\int G_{5X}\sqrt{-X}\, dX \, .
\end{split}
\ee
It is then straightforward to express the above Lagrangians in ADM coordinates (\ref{ADM}).

 \subsubsection{Beyond Horndeski}
 \label{Beyond Horndeski}
 Requiring equations of motion to be at most second order, which leads to Horndeski theories, has long seemed to be a necessary requirement in order to avoid ghost-like instabilities, associated with higher order time derivatives, also known as Ostrogradksi instabilities. However, it has been shown in \cite{GLPV2,GLPV3}  (see also \cite{Lin:2014jga} for similar analysis and conclusion and 
 \cite{Gao:2014soa,Kase:2014yya,Gao:2014fra} for extensions) 
 that an action composed of the Lagrangians 
 \be
\label{Lagrangian_uni}
\begin{split}
 L_2 & \equiv \Atwo(t,N)\;, \\ 
 L_3 &\equiv\Athree (t,N) K\;, \\
L_4 &\equiv \Bfour (t,N) \big(K^2 - K_{ij}K^{ij} \big) + \Afour (t,N) \R\;, \\
L_5 &\equiv \Bfive(t,N) \big(K^3 - 3 K K_{i j}K^{i j} + 2  K_{i j}  K^{i k} K^j_{\ k} \big)  + \Afive (t, N) K^{ij} \bigg( \R_{ij} - \frac12 \gammah_{ij} R \bigg)   \;,
\end{split}
\ee
 with arbitrary functions $B_4$ and $B_5$, i.e.~{\it without assuming} $B_4$ and $B_5$ to depend on, respectively, $A_4$ and $A_5$ (as implied by the Hordenski Lagrangians (\ref{L4ADM}) and (\ref{L5ADM})),\footnote{The Lagrangians \eqref{Lagrangian_uni} describe Horndeski theories if the following relations hold: $A_4 = - B_4 + 2 X B_{4X}$ and $A_5 = -X B_{5X}/3$.}  
 does not lead to Ostrogradski instabilities, in contrast with previous expectations. This conclusion is based on a Hamiltonian analysis of the Lagrangian (\ref{Lagrangian_uni}), which applies to all configurations where the spacetime gradient of the scalar field is timelike. 
 
   Interestingly, one can also map  two subclasses of the general covariant Lagrangian, namely the subclass without $L_4$ and the subclass without $L_5$, to Horndeski theories via a disformal transformation of the metric    (disformal transformations are discussed in section \ref{section_disformal}). Since $L_4$ and $L_5$ require distinct disformal transformations to be related to Horndeski theories,  such transformation cannot be applied to the whole Lagrangian~\cite{GLPV3}.

 \subsubsection{Ho\v{r}ava-Lifshitz theories}
An interesting class of Lorentz-violating gravitational theories has been introduced by Ho\v{r}ava with the goal 
of obtaining
(power counting) renormalizability \cite{Horava:2009uw}. These theories, dubbed Ho\v{r}ava-Lifshitz gravity, assume  the existence of a preferred foliation of spacelike hypersurfaces. An ADM formulation of these theories is thus very natural, even if  a covariant description is also possible, via the introduction of a scalar field, often called ``khronon'', that describes the foliation. 
Several variants of Ho\v{r}ava-Lifshitz gravity have been proposed in the literature. In particular, the so-called healthy non-projectable extension has been shown to be free of instabitilities \cite{Blas:2009yd,Blas:2009qj}. 
All these theories are describable by a Lagrangian of the form (\ref{start_point}), which can be written as (see \cite{Blas:2010hb} for a general discussion)
\be
 \label{HL}
 L_{\text{HL}}=\frac{M_{\text{Pl}}^2 }2\,\left[K_{ij}K^{ij}-\lambda K^2 + \mathcal{V}(\R_{ij},\, N^{-1}\partial_iN)\right]\, .
 \ee
Note that the dependence on $N^{-1}\partial_iN$ has been introduced in the healthy non-projectable extension of Ho\v{r}ava-Lifshitz gravity. Since the Ho\v{r}ava-Lifshitz Lagrangian is already in an ADM form, it is very natural to include these theories in our general approach, as discussed in \cite{Gao:2014soa} (see also \cite{Kase:2014cwa}).

\section{Cosmology: background equations and linear perturbations}
\label{sec:3}
In this section, we analyse from a general perspective  the cosmological dynamics, for the background and linear perturbations, simply starting from a generic Lagrangian of the form (\ref{start_point}).

\subsection{Background evolution}
We first discuss the background equations by considering a spatially flat FLRW spacetime, endowed with the metric 
\be
ds^2=-\bN^2(t) dt^2 +a^2(t) \delta_{ij} dx^i dx^j\,.
\ee
In this spacetime, the intrinsic curvature tensor of the constant time hypersurfaces vanishes, i.e. $\R_{ij}=0$, and the components of the extrinsic curvature tensor are given by 
\be
\label{defH}
K^i_j=H\delta^i_j, \qquad   H\equiv \frac{\dot a}{\bar N a} \,,
\ee
where $H$ is the Hubble parameter.
Substituting into the Lagrangian $L$ of (\ref{start_point}), one thus obtains an homogeneous Lagrangian, which is a function of $\bN(t)$, $a(t)$ and of time:
\be
\bar L(a, \dot a, \bN)\equiv  L\left[K^i_j=\frac{\dot a}{\bN a}\,\delta^i_j, R^i_j=0, N=\bN(t)\right] \;.
\ee

The variation of the homogeneous action, 
\be
\bar{S}_g= \int dt \, d^3x \bN a^3 \bar L,
\ee
leads to 
\be
\delta \bar{S}_g=\int dt d^3x \left\{ a^3\left(\bar L+\bN L_N-3H \F\right)\delta \bN+3a^2 \bN\left(\bar L-3H\F-\frac{\dot \F}{\bN}\right)\delta a\right\}\,,
\ee
where $L_N$ denotes the partial derivative $\partial L/\partial N|_{\rm bgd}$,   evaluated on the homogeneous background. We have also introduced the coefficient $\F$, which is defined from the derivative of the  Lagrangian with respect to the extrinsic curvature, evaluated on the background\footnote{The present formulation of our approach is more general than that given explicitly in \cite{GLPV1}, where we assumed that the Lagrangian $L$ was a function of  specific scalar combinations of the geometric tensors, namely of $K\equiv K^i_i$, $S\equiv K^{ij}K_{ij}$, $R\equiv R^i_i$ and $Z\equiv R_{ij} R^{ij}$. The coefficient $\F$ was then related to the derivatives of $L$ with respect to $K$ and $S$, i.e.~$\F=L_K+2H L_S$. The definition (\ref{F}) enables us to include automatically a dependence on other scalar combinations, such as $K^i_{\, j} K^j_{\, k} K^k_{\, i}$ which appears in $L_5$.}
\be
\label{F}
\left(\frac{\partial L}{\partial K_{ij}}\right)_{\rm bgd}\equiv \F \bar g^{ij}\,,
\ee
where $\bar g^{ij}=a^{-2}\delta^{ij}$ are the spatial components of the inverse background metric. 

If we add some matter minimally coupled to the metric $g_{\mu\nu}$, the variation of the corresponding action with respect to the metric defines the energy-momentum tensor, 
\be
\delta S_{\rm m}=\frac12\int d^4x \sqrt{-g}\,  T^{\mu\nu} \, \delta g_{\mu\nu}\,.
\ee
In a FLRW spacetime, this reduces  to 
\be
\label{linearmat}
\delta \bar S_{\rm m}=\int d^4x  \bN a^3\left(-\rho_{\rm m}\frac{\delta \bN}{\bN}+3 p_{\rm m}\frac{\delta a}{a}\right) \;.
\ee
 Consequently, variation of the total homogeneous action $\bar{S}=\bar{S}_g+\bar{S}_{\rm m}$ with respect to $N$ and $a$ yields, respectively, the first and second Friedmann equations in a very unusual form:
 \be
 \label{Fried1}
 \bar L+\bN L_N-3H \F=\rho_{\rm m}
 \ee
 and 
 \be
 \label{Fried2}
 \bar L-3H\F-\frac{\dot \F}{\bN}=- p_{\rm m}\,.
 \ee
 These two equations also imply
 \be
 \frac{\dot \F}{\bN}+\bN L_N=\rho_{\rm m}+p_{\rm m}\,.
 \ee
 
 Although written in a very unusual form, it is easy to check that one recovers the usual Friedmann equations when gravity is described by GR. Indeed, in this case,
 \be 
 \frac{\partial L_{\rm GR}}{\partial K^i_j}=\Mp^2\left(K^j_i-K \d^j_i\right),
 \ee
 which, after substituting $K^i_j=H \d^i_j$, yields, 
 \be
 \F_{\rm GR}=-2 \Mp^2 H\,,
 \ee
 whereas $\bar{L}_{\rm GR}=-3 \Mp^2 H^2$ and $L_N=0$.

\subsection{Quadratic action}
\label{sec:QA}
 In order to describe the dynamics of linear perturbations about the FLRW background solution, we now expand the action up to quadratic order. The tensor $R_{ij}$ vanishes in the background and is thus a perturbative quantity. It is useful to introduce the two other perturbative quantities (remembering the definition of $H$ in eq.~\eqref{defH})
\be
\delta N\equiv N-\bar N\,,\qquad \delta K_i^j\equiv K_i^j- {H} \delta_i^j\,.
\ee
The expansion of the Lagrangian $L$ up to quadratic order yields
\be
\label{L_up_to_2}
L(N, K^i_j, \R^i_j,\dots)=\bar L+L_N\d N+\frac{\partial L}{\partial K^i_j} \d K^i_j+\frac{\partial L}{\partial R^i_j} \d R^i_j+
L^{(2)}+\dots ,
\ee
with the quadratic part given by
\be
\label{L_quad}
\begin{split}
L^{(2)}
  = & \frac12 L_{NN} \d N^2+\frac12\frac{\partial^2 L}{\partial K^i_j\partial K^k_l} \d K^i_j \d K^k_l+\frac12\frac{\partial^2 L}{\partial R^i_j\partial R^k_l} \d R^i_j \d R^k_l+
 \cr
 & + \frac{\partial^2 L}{\partial K^i_j\, \partial R^k_l}\d K^i_j \d R^k_l+
 \frac{\partial^2 L}{\partial N\partial K^i_j} \d N\d K^i_j+\frac{\partial^2 L}{\partial N \partial R^i_j} \d N\d R^i_j+\dots \;,
 \end{split}
 \ee
 where all the partial derivatives are evaluated on the FLRW background (without explicit notation, as will be the case in the rest of this paper). The coefficient $L_{NN}$ denotes the second derivative of the Lagrangian with respect to $N$. The dots in the two above equations correspond to other possible terms which are not indicated explicitly to avoid too lengthy equations, but can be treated exactly in the same way. This includes for instance the spatial derivatives of the curvature or of the lapse, which appear in Horava-Lifshitz gravity. 
 
 The third term on the right hand side of (\ref{L_up_to_2}) can  be simplified as follows. 
 Rewriting it as  
 \be
 \frac{\partial L}{\partial K^i_j} \d K^i_j=\F \d K=\F (K-3H)\,,
 \ee
 and noting that $K=\nabla_\mu n^\mu$, 
 one can  use the integration by parts
\beq
\int d^4x \sqrt{-g} \, \F K=-\int d^4x \sqrt{-g} \, n^\mu \nabla_\mu {\cal F}=-\int d^4x \sqrt{-g} \frac{\dot {\cal F}}{N}\,.
\eeq
This implies that the   Lagrangian (\ref{L_up_to_2})  can be replaced by the equivalent Lagrangian 
\be
L^{\rm new}= \bar L -3H\F-\frac{\dot \F}{N}+L_N\d N+ \frac{\partial L}{\partial R^i_j} \delta R^i_j+ L^{(2)} \;.
 \ee

Let us now consider the quadratic part (\ref{L_quad}). Because of the background geometry, the coefficient of the second term is necessarily of the form\footnote{This is equivalent to  the definition below, expressed with covariant indices for the extrinsic curvature  tensors, which makes the symmetry under exchange of the indices more manifest:
$$
\frac{\partial^2 L}{\partial K_{ij}\,\partial K_{kl}}\equiv \A\,  \bar g^{ij} \, \bar g^{kl}+ \LS\left( \bar g^{ik} \, \bar g^{jl}+\bar g^{il} \, \bar g^{jk}\right)\,.
$$ 
} 
\be
\label{AAA}
\frac{\partial^2 L}{\partial K_i^j\,\partial K_k^l}=\A \,  \d^i_j \, \d^k_l+ \LS\left(\d^i_l\, \d^k_j+\d^{ik} \d_{jl}\right)\,,
\ee
where we have introduced  the (a priori time-dependent) coefficients $\A$ and $\LS$. Similarly, one can write
\be
\frac{\partial^2 L}{\partial R_i^j\, \partial R_k^l}=\LRR\,  \d^i_j \, \d^k_l+ \LZ \left(\d^i_l\, \d^k_j+\d^{ik} \d_{jl}\right)\,,
\ee
and 
\be
\frac{\partial^2 L}{\partial K_i^j\, \partial R_k^l}= \Co\,  \d^i_j \, \d^k_l+ \Ch \left(\d^i_l\, \d^k_j+\d^{ik} \d_{jl}\right)\,.
\ee
The mixed coefficients that appear on the second line of eq.~\eqref{L_quad} are proportional to $\delta_i^j$ and can be written 
as
\be
\frac{\partial^2 L}{\partial N\partial K^i_j} =\B \, \d_i^j\,, \qquad 
\frac{\partial^2 L}{\partial N\partial R^i_j}={\LNRo} \, \d_i^j\,. \label{BBB}
\ee

Taking into account the term $\sqrt{-g}=N\sqrt{h}$, it is straightforward to derive the quadratic part of the full Lagrangian ${\cal L}\equiv \sqrt{-g} \, L$, which is relevant to study linear perturbations. 
After some cancellations due to the background equations of motion\footnote{If matter is present,  one must also include in the quadratic Lagrangian the terms from the expansion of the matter action with respect to the metric perturbations.}, one  finds
\be
\label{L2new}
\begin{split}
{\cal L}_2&=    \bar{N} \LRo\, \d_1 \sR\, \d \sqrt{h}+ a^3\left(L_N+\frac{1}2\bar N L_{NN}\right)\d N^2
\cr
&
+ \bar{N} a^3\left[\LRo \d_2\sR +\frac12\A \, \d K^2 
+\B \, \d K \d N
+
\Co\, \d  K \d \sR +\Ch\, \d K^i_j \, \d R^j_i
\right.
\cr
& \left.
+ \LS \, \d K^i_j \, \d K^j_i
+\LZ \,\d \sR^i_j \, \d \sR^j_i
+ \frac12  \LRR \; \d \sR^2
+ \left(\frac{\LRo}{\bar N} + {\LNRo}\right) \, \d N \d\sR
\right]+\dots \;,
\end{split}
\ee
where, in analogy with the definition (\ref{F}) of $\F$, we have introduced the coefficient $\LRo$ defined by 
\be
\frac{\partial L}{\partial R^i_j}=\LRo \, \d_i^j\,.
\ee 
We have also denoted 
as
$\d_1 R$ and $\d_2 R$, respectively, the first  and second order terms of the curvature $\R$ expressed in terms of the metric perturbations.

Note that the coefficients that enter here in the quadratic Lagrangian are more general than those introduced explicitly in \cite{GLPV1}, where the  Lagrangian $L$ was considered as a function of $N$, 
$K$, $S= K_{ij}K^{ij}$, $R$ and $Z\equiv R_{ij}R^{ij}$. It is however straightforward to derive the relation between the present coefficients in terms of our former notation\footnote{
The correspondence is given by $\A= 4H^2 L_{SS}+4HL_{SK}+L_{KK}$, $\LS=L_S$, $\B=2H L_{SN}+L_{KN}$, $\LNRo=L_{NR}$, $\LZ=L_Z$, $\LRo=L_R$, $\LRR=L_{RR}$ and $\Co=2H L_{SR}+L_{KR}$. 
\label{footnote_L}
}. The present definitions have the advantage to 
automatically include
cases with more complicated combinations involving the tensors $K_{ij}$ or $R_{ij}$, such as $K_i^j K_j^k K^i_k$ in the Lagrangian term $L_5$  that appears in  Horndeski  theories and beyond.

The above quadratic expression can be further simplified, as shown in \cite{GLPV1}, by reexpressing $\d K^i_j \, \d R^j_i$ in terms of the other terms, thanks to the identity
\be
\label{app_RK}
\int d^4 x \sqrt{-g}\, \lambda(t) R_{ij}K^{ij}  =  \int d^4 x \sqrt{-g} \left[ \frac{\lambda(t)}{2} R\; K + \frac{\dot \lambda(t)}{2 N} \; R \right]\;.
\ee
This implies  the following replacement at quadratic order:
\be
\bar{N} a^3 \Ch\, \d K^i_j \, \d R^j_i \quad \rightarrow \quad \frac{\bar{N} a^3}{2}\left[ \left(\frac{\dot\Ch}{\bar{N}}+H\Ch\right)\left(\d_2 R+\frac{\d\sqrt{h}}{a^3}\d R\right)+ \Ch \, \d R\d K + \frac{H\Ch}{\bar{N}}\d N \d R\right]\,.
\ee
Consequently, the quadratic Lagrangian (\ref{L2new}) is equivalent to the new one
\be
\label{L2new2}
\begin{split}
{\cal L}_2^{\rm new} &=    \bar{N} \LR\, \d_1 \sR\, \d \sqrt{h}+ a^3\left(L_N+\frac12 \bar N L_{NN}\right)\d N^2
\cr
&
+ \bar{N} a^3\left[\LR \d_2\sR +\frac12\A \, \d K^2 
+\B \, \d K \d N
+
\C\, \d  K \d \sR 
\right.
\cr
& \left.
+ \LS \, \d K^i_j \, \d K^j_i
+\LZ \,\d \sR^i_j \, \d \sR^j_i
+ \frac12  \LRR \; \d \sR^2
+ \left(\frac{\LR}{\bar N} + \LNR\right) \, \d N \d\sR
\right]+\dots \;,
\end{split}
\ee
with the ``renormalized'' coefficients
\footnote{For a Lagrangian $L$ which is a function of $N$, 
$K$, $S= K_{ij}K^{ij}$, $R$, $Z\equiv R_{ij}R^{ij}$ and {\em also} of $Y \equiv R_{ij} K^{ij}$, the relation between the coefficients defined in this paper and the derivatives of $L$ with respect to the above quantities is unchanged for $\A$, $\LS$, $\B$ and $\LZ$ (see footnote \ref{footnote_L}). The other coefficients, taking into account the dependence on $Y$,  are given by $\LNRo^* = L_{NR}^* \equiv L_{NR} + H L_{NY} - \dot L_Y/2$, $\LRo^* = L_R^* \equiv L_R + \dot L_Y/2 + 3 H L_Y/2$, $\LRR = L_{RR} + H^2 L_{YY} + 2 HL_{YR}$ and $\Co^* =  2H L_{SR}+L_{KR}  + HL_{KY} + 2 H^2 L_{SY} + L_Y/2$ with $\bar N=1$
.}
\be
\begin{split}
&
\LR=\LRo+\frac{\dot\Ch}{2\bar{N}}+H\Ch\,,
\cr
& 
\C=\Co+\frac12 \Ch\,,
\cr
&
\LNR={\LNRo}-\frac{\dot\Ch}{2\bar{N}^2}\,.
\end{split}
\ee

\subsubsection{Tensor modes}
Let us first investigate the tensor modes in the general  quadratic Lagrangian (\ref{L2new2}). 
At linear order, tensor modes correspond to the perturbations of  the spatial metric
\be
h_{ij} = a^2(t) \left(\delta_{ij} + \gamma_{ij}\right) \;,
\ee
with $\gamma_{ij}$ traceless and divergence-free,  $\gamma_{ii}=0 = \partial_i \gamma_{ij}$. 
Using 
\be
\d K^i_j=\frac{1}{2\bar N}\dot\gamma^i_{\, j}
\ee
and
\be
\d_2\R=\frac{1}{a^2}\left(\gamma^{ij}\partial^2 \gamma_{ij}+\frac34 \partial_k \gamma_{ij}\partial^k \gamma^{ij}-\frac12 \partial_k \gamma_{ij} \partial^j \gamma^{ik}\right)\;,
\ee
one finally obtains
\be
S_{\gamma}^{(2)} =\int d^3x dt \, a^3  \left[ \frac{\LS}{4} \dot{\gamma}_{ij}^2 -\frac{\LR}{4a^2}(\partial_k \gamma_{ij})^2 \right] \;, \label{gamma_action}
\ee
where here and below we set $\bar N =1$.
We recover the standard GR result when $\LS=\LR=M_{\rm Pl}^2 /2$. By comparison, this suggests to define the effective Planck mass squared by 
\be
M^2 \equiv 2 \LS > 0\,,
\ee
where the sign is required to avoid ghost instabilities,
and write the action as
\be
S_{\gamma}^{(2)} =\int d^3x dt \, a^3  \frac{M^2}{8} \left[  \dot{\gamma}_{ij}^2 -\frac{c_T^2}{a^2}(\partial_k \gamma_{ij})^2 \right] \;.
\ee
The square of the graviton propagation speed is  given by 
\be
\label{alphaT}
c_T^2\equiv 1+\alpha_T= \frac{\LR}{\LS}\,,
\ee
where $\alpha_T$ represents the deviation with respect to the GR result.

The graviton sector is thus characterized by the two coefficients $\LS$ and $\LR$, or equivalently by 
$M$ and $\alpha_T$. In practice, it is the time variation which can distinguish the effective Planck mass defined here with respect to the standard Planck mass, so it is convenient, 
following \cite{Bellini:2014fua},
to introduce the dimensionless parameter 
\be
\label{alphaM}
\alpha_M \equiv 
\frac1H
\frac{d}{dt}\ln M^2\,.
\ee
With these definitions, the evolution equation for tensor modes is given by
\be
\ddot \gamma_{ij} + H(3+\alpha_M) \dot \gamma_{ij} - (1+\alpha_T) \frac{\nabla^2}{a^2} \gamma_{ij} = \frac{2}{M^2} 
\left(T_{ij} -  \frac{1}3 T \delta_{ij} \right)^{TT} \;,
\ee
where $(T_{ij} -  T \delta_{ij}/3)^{TT}$ is the transverse-traceless projection of the anisotropic matter stress tensor.

\subsubsection{Vector modes}

Let us now study the behaviour of vector modes. In unitary gauge, these are parameterized by the transverse components of the shift vector, i.e.
$N^i = N^i_V$ with $\partial_i N^i_V =0$. The second-order action for vector modes then is
\be
S^{(2)}_{N_V} = \int d^3x dt \, \frac{1}{a}  \frac{M^2}{8}  (\partial_i N^V_j + \partial_j N^V_i)^2 \;.
\ee
Including matter, variation of the action with respect to $N_i^V$ gives the transverse part of the momentum constraint,
\be
\frac12 \nabla^2 N^V_i = \frac{a^2}{M^2} \big( T^0_{\ i}\big)^T \;,
\ee
where $( T^0_{\ i})^T$ is the transverse projection of the matter energy flux. For a perfect fluid, the conservation of the matter stress-energy tensor implies that
$( T^0_{\ i})^T \propto 1/a^3$; then the metric vector perturbations scale as
\be
N^i_V \propto \frac{1}{a M^2} = \frac{1}{a^{1+\alpha_M}} \;,
\ee
where the last equality holds for a constant $\alpha_M$. 
It is interesting to note that the evolution of the vector modes  is modified when the gravitational effective mass $M$ is time-dependent, i.e. when $\alpha_M\neq 0$. Thus, in principle, measuring the time evolution of the  vector and tensor perturbations could determine $\alpha_M$ and $\alpha_T$,  independently of the  scalar modes.

\subsubsection{Scalar modes}
Without loss of generality, 
in unitary gauge
the scalar modes can be described 
by the 
metric
perturbations  
\cite{Maldacena:2002vr}
\be
N=1+\d N, \quad N^i=\delta^{ij}\partial_j \psi, \quad {h}_{ij}=a^2(t) e^{2\zeta}\delta_{ij}\,. \label{metric_ADM_pert}
\ee
Substituting 
\be
\label{dhdK}
\d\sqrt{h} = 3 a^3 \zeta\,, \qquad  \d K^i_{\ j}=\left(\dot\zeta-H\d N\right)\d^i_j-\frac{1}{a^2}\d^{ik}\partial_{k}\partial_j\psi \;,
\ee
and
\be \label{RRR}
\d_1\sR_{ij}
=  - \delta_{ij} \partial^2 \zeta -  \partial_i \partial_j \zeta \;, \qquad \d_2 \sR=  -\frac{2}{a^2}\left[(\partial\zeta)^2-4\zeta\partial^2\zeta\right]\,,
\ee
into  (\ref{L2new2}), one obtains a lengthy Lagrangian in terms of $\delta N$, $\psi$ and $\zeta$. Since the Lagrangian does not depend on the time derivatives of the lapse and of the shift, the variation of the Lagrangian with respect to  $\delta N$ and $\psi$ yields two constraints, corresponding to the familiar Hamiltonian constraint and (the scalar part of) the momentum constraint. 

In the following, we will assume for simplicity  the conditions 
\beq
\label{conds}
\A+2\LS =0\;, \qquad \C =0 \; ,\qquad 4\LRR+3\LZ=0\; ,
\eeq
which ensure that there are at most two spatial derivatives in the quadratic Lagrangian written in terms of $\zeta$ only. This includes Horndeski theories as well as their extensions discussed in Section~\ref{Beyond Horndeski}.
 
Provided conditions (\ref{conds}) are satisfied, one finds that the momentum constraint reduces to
\be
\label{soldn}
\d N=\frac{4 \LS}{\B+4 H\LS} \, \dot\zeta= \frac{\dot\zeta}{H \left(1+\aB \right)}\,,
\ee
where we have introduced the dimensionless 
quantity\footnote{Although we use the same symbol, our variable $\aB$ differs from that introduced in \cite{Bellini:2014fua} by a factor $-2$. This simplifies the subsequent equations.}
\be
\label{alphaB}
\aB\equiv \frac{\B}{4H \LS}\,,
\ee
which expresses the deviation from the standard expression
$\d N=\dot\zeta/H$.
When $\alpha_B\neq 0$, part of the kinetic term  of scalar fluctuations comes from the term $\delta K \delta N$ in action \eqref{L2new2}, i.e. from kinetic mixing between gravitational and scalar degrees of freedom \cite{Creminelli:2006xe,Cheung:2007st,Creminelli:2008wc}. This phenomenon has been called {\em kinetic braiding} in \cite{Deffayet:2010qz,Pujolas:2011he}.

Substituting~\eqref{soldn}, the quadratic action for $\zeta$ is then given by
\be 
\label{lag-quad}
S^{(2)}= \frac12 \int d^3x dt \, a^3  \bigg[ {\cal L}_{\dot \zeta \dot \zeta} \dot \zeta^2  +  {\cal L}_{\partial \zeta \partial \zeta} \frac{(\partial_i \zeta)^2}{a^2}    + \frac{M^2}{4} {\dot \gamma}_{ij}^2 - \frac{M^2}{4}  (1+ \alpha_T) \frac{(\partial_k \gamma_{ij})^2}{a^2}   \bigg]\;,
\ee
with 
\begin{align}
{\cal L}_{\dot \zeta \dot \zeta} &\equiv   M^2 \,  \frac{\DD}{(1+ \aB)^2}  \;, \qquad \DD \equiv \alpha_K + 6 \alpha_B^2 \;, \\
{\cal L}_{\partial \zeta \partial \zeta} &\equiv 2  M^2 \left\{1+\alpha_T -\frac{1+ \alpha_H}{1+ \aB} \Big(1+\alpha_M-\frac{\dot H}{H^2}\Big)
-\frac{1}{H}\frac{d}{dt} \left(\frac{1+ \alpha_H}{1+ \aB}\right)\right\}
 \;, 
 \end{align}
where we have introduced the dimensionless time-dependent functions 
\be
\label{alphaKH}
\alpha_K\equiv \frac{2L_N+L_{NN}}{2H^2 \LS}\,, \qquad 
\alpha_H \equiv \frac{\LR+\LNR}{\LS}-1\,.
\ee
Note that the  coefficient of the kinetic term reduces to ${\cal L}_{\dot \zeta \dot \zeta}=M^2\alpha_K$ when $\alpha_B=0$. 
In this case, the kinetic  coefficient for $\zeta$ is directly related  to the coefficient of  the term $\delta N^2$ in the quadratic Lagrangian  (\ref{L2new2}), which represents the kinetic energy of the scalar field fluctuations. 
The parameter $\alpha_H$ is different from zero for theories that  deviate from Horndeski theories \cite{GLPV1,GLPV2,GLPV3}.  In particular, this includes theories that can be related to Horndeski theories via disformal transformations, as shown in  \cite{GLPV3}. Indeed, starting from a Horndeski theory  for a metric $\tilde g_{\mu\nu}$ related to $g_{\mu\nu}$  via a disformal transformation that depends on $X$, the Lagrangian expressed in terms of $g_{\mu\nu}$ differs from the standard Horndeski Lagrangian, which implies $\alpha_H\neq 0$.

Classical and quantum stability (absence of ghosts) requires 
the kinetic coefficient to be  positive, 
\be
\label{stability}
{\cal L}_{\dot \zeta \dot \zeta}>0\quad \implies \quad \DD = \alpha_K + 6 \alpha_B^2 > 0\;.
\ee
The sound speed (squared) of fluctuations can be simply computed by taking the ratio
\be
\label{sound_speed}
c_s^2 \equiv -\frac{ {\cal L}_{\partial\zeta \partial \zeta} }{ {\cal L}_{\dot \zeta \dot \zeta} }\;.
\ee
When adding matter to the dark energy Lagrangian, 
the kinetic and spatial gradient terms of the scalar fluctuations acquire  new contributions that modify the expression for the sound speed \cite{GLPV2,GLPV3}.
The final expression for the sound speed,  when matter is present, reads
\beq
c_s^2 =   -2 \frac{(1+\alpha_B)^2}{\DD} 
\left\{1+\alpha_T -\frac{1+ \alpha_H}{1+ \aB} \Big(1+\alpha_M-\frac{\dot H}{H^2}\Big)
-\frac{1}{H}\frac{d}{dt} \left(\frac{1+ \alpha_H}{1+ \aB}\right)\right\}
- \frac{(1+\alpha_H)^2}{\DD}\, \frac{\rho_{\rm m} + p_{\rm m}}{ M^2H^2}\;.
\eeq
In the simple case of $k$-essence field with a Lagrangian $P(\phi,X)$, where all $\alpha_i$ coefficients vanish except $\alpha_K=(2\bar  X P_{X}+ 4 \bar X^2 P_{XX})/(M^2 H^2)$, the above formula yields 
$c_s^2= -2 \dot H/(\alpha_K H^2)-  (\rho_{\rm m}+p_{\rm m})/ (\alpha_K M^2 H^2)$ 
and one recovers $c_s^2=P_{X}/(P_{X}+ 2 \bar X P_{XX})$ after using the Friedmann equation $\dot H=- ( 2\bar  X P_{X} +\rho_{\rm m}+p_{\rm m})/(2M^2)$.

\subsection{Link with the building blocks of dark energy}
In the previous subsection, we have focussed our attention on Lagrangians that satisfy the conditions (\ref{conds}) in order to get propagation equations with no more than two (space) derivatives. 
At  quadratic order, the most general action of the form~\eqref{L2new2} that satisfies these conditions can be written in the form
\be
\begin{split}
\label{SBAction}
S^{(2)} = \ \int d^3x dt a^3\, \frac{M^2}2\bigg[ &\delta K_{ij }\delta K^{ij}-\delta K^2  +(1+\alpha_T) \bigg( \R \frac{\delta \sqrt{h}}{a^3} + \delta_2 R \bigg)\\
&  + \alpha_K H^2 \delta N^2+4\alpha_B H \, \delta K\, \delta N+ ({1+\alpha_H}) \R\, \delta N \bigg]\, ,
\end{split}
\ee
where, for convenience, we summarize in Table~\ref{table} the definitions of the parameters $\alpha_i$ introduced 
in the previous subsection, in terms of the original coefficients  defined in Sec.~\ref{sec:QA} (second row) and those introduced explicitly in Ref.~\cite{GLPV1} (third row). 

\renewcommand{\arraystretch}{2.2}
\begin{table}
\begin{center}
\begin{adjustbox}{max width=\textwidth}
  \begin{tabular}{ | l | c | c | c | c | c | c | c | }
    \hline
    Eq.~\eqref{SBAction}  & $M^2$ & $\alpha_M$ & $\alpha_K$ &$ \alpha_B $& $\alpha_T $& $\alpha_H$ \\  \hline   
     Eq.~\eqref{L2new2}       & $\displaystyle 2 \LS $ & $\displaystyle \frac1H\frac{d}{dt}\ln  \LS $ & $\displaystyle \frac{2L_N+L_{NN}}{2H^2 \LS}$ &$\displaystyle \frac{\B}{4H \LS} $& $\displaystyle \frac{\LR}{\LS} -1 $& $\displaystyle \frac{\LR+\LNR}{\LS}-1 $ \\ [0.2cm] \hline
          Eq.~(12) of \cite{GLPV1}       & $\displaystyle 2 L_S $ & $\displaystyle \frac1H\frac{d}{dt}\ln  L_S  $ & $\displaystyle \frac{2L_N+L_{NN}}{2H^2 L_S }$ &$\displaystyle \frac{2 H L_{S N} + L_{KN} }{4H L_S } $& $\displaystyle \frac{L_R^*}{L_S} -1 $& $\displaystyle \frac{L_{R}^*+L_{NR}^*}{L_S }-1 $ \\ [0.2cm] \hline
      Eq.~\eqref{Lagrangian_uni}         Eq.~\eqref{total_action}     & $\MM^2f+2\mfs$ & $\displaystyle \frac{\MM^2 \dot f+2 (\mfs)^{\hbox{$\cdot$}}}{M^2 H}$ & $\displaystyle\frac{2c+4M_2^4}{M^2H^2} $ &$\displaystyle \frac{\MM^2 \dot f-m_3^3}{2M^2H} $& $\displaystyle-\frac{2\mfs}{M^2}$& $\displaystyle\frac{2(\tmfs-\mfs)}{M^2}$   \\  [0.1cm]  \hline
  \end{tabular}
\end{adjustbox}
\end{center}
\caption{In the first row, the parameters $\alpha_i$ introduced in eqs.~\eqref{alphaT}, \eqref{alphaM}, \eqref{alphaB} and \eqref{alphaKH}, i.e.~the Lagrangian coefficients of eq.~\eqref{SBAction}. These parameters are written in terms of the Lagrangian coefficients of eq.~\eqref{L2new2}, defined in eqs.~\eqref{AAA}--\eqref{BBB} (second row), of  the coefficients  introduced  in \cite{GLPV1}, where the derivative of the  Lagrangian $L$ with respect to $N$, 
$K$, $S= K_{ij}K^{ij}$, $R$, $Z\equiv R_{ij}R^{ij}$ and $Y \equiv R_{ij} K^{ij}$ (third row) and, finally, of the EFT Lagrangian, action \eqref{total_action} (fourth row). All these quantities are understood to be evaluated on the background, with $\bN =1$.}
  \label{table}
\end{table}

The action leading to the quadratic Lagrangian (\ref{SBAction}) can also be written in the standard EFT form, 
with an explicit dependence on the four-dimensional scalar curvature,  $g^{00}$ and several quadratic operators. 
This action  reads \cite{GLPV1}
\be
\begin{split}
\label{total_action}
S =& \int \! d^4x \sqrt{-g} \left[\, \frac{\MM^2}{2} f(t) {}^{(4)}\!R - \Lambda(t) - c(t) g^{00}  + \, \frac{M_2^4(t)}{2} (\delta g^{00})^2\, -\, \frac{m_3^3(t)}{2} \, \delta K \delta g^{00} \, 
 \right. \\[1.2mm]
 &  - \left. \,  m_4^2(t)\left(\delta K^2 - \d K^\mu_{ \ \nu} \, \d K^\nu_{ \ \mu} \right) \, +\, \frac{\tilde m_4^2(t)}{2} \, \R \, \delta g^{00}  \right] \;.
\end{split}
\ee
It leads to the background equations of motion \cite{GLPV1}
\begin{align}
c+ \Lambda &= 3 \MM^2 (f H^2 + \dot fH) -\rho_{\rm m} \;, \label{background1} \\
\Lambda - c &= \MM^2 (2 f \dot H + 3 f H^2 + 2 \dot f H + \ddot f) + p_{\rm m} \label{background2} \;,
\end{align}
and to the quadratic action  for the linear perturbations \eqref{SBAction}, where the relation between the coefficients 
$\alpha_i$ and the seven parameters appearing in   (\ref{total_action}) is given in Table~\ref{table}.   
The two background equations of motion \eqref{background1} and \eqref{background2} imply that  only five of the EFT parameters are independent, 
thus setting the minimal number of functions parametrizing deviations from General Relativity \cite{GLPV1}.

\subsection{Disformal transformations and dependence on $\dot N$}
\label{section_disformal}
In our discussion, 
we have assumed that the initial Lagrangian depends on $N$, but not on its time derivative $\dot N$. Allowing a dependence on $\dot N$ leads in general to an additional propagating degree of freedom. However, this is not always the case, as illustrated by considering disformal transformations of the metric,
 originally introduced in \cite{Bekenstein:1992pj}, 
 of the form 
\begin{equation} \label{disformal}
g_{\mu \nu} \to \tilde g_{\mu \nu} = \Omega^2(\phi,X)\,  g_{\mu \nu} + \ga(\phi,X)\,  \partial_\mu  \phi \, \partial_\nu \phi \, .
\end{equation}
As shown in \cite{Bettoni:2013diz}, Horneski theories are invariant under a restricted class of disformal transformations where $\Omega$ and $\Gamma$ depend on $\phi$ only, not on $X$. In \cite{GLPV3}, we showed explicitly that one could use disformal transformations with an $X$-dependent function $\Gamma$ to relate 
subclasses of theories beyond Horndeski
to Horndeski theories. A similar result for a disformal transformation of the Einstein-Hilbert Lagrangian was 
previously obtained
in \cite{Zumalacarregui:2013pma}.

In unitary gauge, $\Omega$ and $\Gamma$ become functions of  the time variable $t$ and of the lapse function $N$. By  choosing time to coincide with $\phi$, i.e. $\partial_\mu \phi = \delta_\mu^0$, the disformal transformation \eqref{disformal}  corresponds, in the ADM language, to the transformations 
\cite{GLPV3}
\begin{equation} 
\label{3-d}
\tilde N^i  = N^i \;, \qquad 
\tilde h_{ij} = \Omega^2 (t,N) \, h_{ij} \;, \qquad 
\tilde N^2= \Omega^2(t,N) \, N^2 - \ga \, (t,N)\, . 
\end{equation}
Moreover, the relations between the old and new curvature tensors are given by
\begin{equation} \label{mammate}
\tilde\R = \Omega^{-2}\left[\R - 4 D^2 \ln \Omega - 2 \partial_i (\ln \Omega) \partial^i (\ln \Omega)\right]\, ,
\end{equation} 
and 
\begin{equation}
\tilde K_{ \ i}^j \ =\ \frac{N}{\tilde N} \left[K^j_{\ i} \, - \,N  g^{0 \mu} \partial_\mu \ln \Omega \, \delta_{ \ i}^j \right]. 
\end{equation}
The last relation can be expanded into
\be
\label{tilde_K}
\tilde K_{ \ i}^j \ =\ \frac{N}{\tilde N} \left[K^j_{\ i} \, +\frac{1}{N \Omega}\left(\Omega_t +\Omega_N\big(\dot N -N^i\partial_i N\big)\right) \, \delta_{ \ i}^j \right]. 
\ee
Consequently, a Lagrangian that depends initially on tilded quantities, will finally depend on $\dot N$ when reexpressed in terms of untilded quantities.
The quadratic Lagrangian will now depend on $\delta \dot N$, in addition to all the terms discussed previously. However, according to  (\ref{dhdK}) and (\ref{tilde_K}), one sees that $\delta \dot N$ will always appear associated with $\dot \zeta$ in the combination 
\be
\dot\zeta+\frac{\Omega_N}{\tilde N \Omega}\d \dot N\,,
\ee
which implies that the matrix of the kinetic coefficients is degenerate. Thus,  one can introduce a new degree of freedom
\be
\zeta_{\rm new}=\zeta+\frac{\Omega_N}{\tilde N \Omega}\delta N\,,
\ee
which absorbs all time derivatives of $\delta N$. Contrarily to what  could have been expected, the explicit dependence on $\dot N$ does not lead, in this particular case, to an extra degree of freedom.  

\section{Evolution of the cosmological perturbations}
\label{sec:perts}

In this section we follow \cite{GLPV1} and derive the evolution equations for linear scalar perturbations described by the action \eqref{SBAction},
together with some matter field minimally coupled to the metric $g_{\mu \nu}$.
We first restore the general covariance of the action 
and write it in a generic coordinate system. In order to do so, we perform the time diffeomorphism \cite{ArkaniHamed:2003uy,Creminelli:2006xe,Cheung:2007st}
\beq
t \to t + \pi (t, \vec x) \;,
\eeq
where $\pi$ describes the  fluctuations of the scalar degree of freedom.
Under this time diffeomorphism, any function of time $f$  changes up to second order as 
\begin{align}
f &\to f + \dot f \pi  + \frac12 \ddot f \pi^2 + {\cal O} (\pi^3)\;,  \label{trans_ST_6} 
\end{align}
while the metric component $g^{00}= -1/N^2$ 
exactly
transforms as
\beq
g^{00} \to g^{00} +  2 g^{0 \mu} \partial_\mu \pi + g^{\mu \nu} \partial_\mu \pi \partial_\nu \pi \;.
\eeq
For the other perturbed geometric quantities, we only need their change at linear order in $\pi$, i.e. \cite{GLPV1}
\begin{align}
\delta K_{ij} &\to \delta K_{ij} - \dot H \pi h_{ij} - \partial_i \partial_j \pi  + {\cal O} (\pi^2) \;, \\
\delta K &\to \delta K  - 3 \dot H \pi - \frac1{a^2} \partial^2 \pi + {\cal O} (\pi^2) \;, \\ 
R_{ij} &\to R_{ij}  + H (\partial_i \partial_j \pi + \delta_{ij} \partial^2 \pi) + {\cal O} (\pi^2) \;, \\
R &\to R + \frac4{a^2} H \partial^2 \pi + {\cal O} (\pi^2) \;. \label{variationR}
\end{align}
We stress that in the above expressions $K_{ij}$ and $R_{ij}$ respectively denote the extrinsic and intrinsic curvature on hypersurfaces of constant time, even when we are {\it not} in unitary gauge. Therefore they are not the same geometrical quantities {\it before} and {\it after} the change of time. 

We can then expand the covariant action  up to quadratic order, considering a linearly perturbed FLRW metric. 
Varying the action with respect to the four scalar perturbations in the metric and the scalar fluctuation $\pi$ we obtain five scalar equations; see Ref.~\cite{GLPV1} for details on their derivations. We turn to a discussion of these equations restricting to Newtonian gauge.

\subsection{Perturbation equations in Newtonian gauge}
\label{sec:Einstein}

We assume a perturbed FLRW metric in Newtonian gauge with only scalar perturbations, i.e.,
\be
\label{metric_Newtonian}
ds^2 = - (1+2 \Phi) dt^2  + a^2(t) (1-2 \Psi) \delta_{ij}   dx^i dx^j \;.
\ee
The metric perturbations $\Phi$ and $\Psi$ and the scalar fluctuation $\pi$ are related to the metric perturbations in unitary gauge defined in eq.~\eqref{metric_ADM_pert} by
\be
\Phi = \delta N + (a^2 \psi)^{\hbox{$\cdot$}} \;, \qquad \Psi = - \zeta - a^2 H \psi \;, \qquad \pi = a^2 \psi \;.
\ee
Moreover, in this gauge we decompose the 
total
matter stress-energy tensor at linear order  as 
\begin{align}
T^0_{\ 0} &\equiv - (\rho_{\rm m} + \delta \rho_{\rm m}) \;, \label{se1}\\
T^0_{\ i} &\equiv \partial_i q_{\rm m}  \equiv (\rho_{\rm m} + p_{\rm m}) \partial_i v_{\rm m} = - a^2 T^i_{\ 0}\;, \label{se2} \\
T^i_{\ j} &\equiv (p_{\rm m} + \delta p_{\rm m}) \delta^i_j + \left( \partial^i \partial_j - \frac13 \delta^i_j \partial^2 \right) \sigma_{\rm m} \label{se3}\;,
\end{align}
where  $\delta \rho_{\rm m} $ and $\delta p_{\rm m}$ are the  energy density and pressure perturbations, $q_{\rm m}$ and $v_{\rm m}$ are respectively the 3-momentum and the 3-velocity potentials;  $\sigma_{\rm m}$ is the  
anisotropic stress potential.

The Hamiltonian constraint ((00) component of the Einstein equation) is
\be
\begin{split}
&6 (1+\alpha_B)H \dot \Psi + (6-\alpha_K +12 \alpha_B )H^2  \Phi + 2 (1+ \alpha_H) \frac{k^2}{a^2}\Psi \\
& +  \left(\alpha _K -6 \alpha_B \right)H^2 \dot \pi +6 \left[  (1+\alpha_B) \dot H + \frac{\rho_{\rm m} + p_{\rm m}}{2 M^2}   + \frac13 \frac{k^2}{a^2} (\alpha_H -\alpha_B) \right]  H  \pi = -\frac{\delta \rho_{\rm m}}{ M^2} \;, \\ \label{Ein1}
\end{split}
\ee
while the momentum constraint ((0$i$) components of the Einstein equation) reads
\be
2 \dot \Psi + 2(1+\alpha_B) H \Phi -2 H \alpha_B \dot \pi +  \bigg( 2\dot H  +\frac{\rho_{\rm m} + p_{\rm m}}{ M^2} \bigg) \pi= - \frac{(\rho_{\rm m} + p_{\rm m}) v_{\rm m}}{M^2}  \;.\label{Ein2}
\ee
The traceless part of the $ij$ components of the Einstein equation gives
\be
 (1+ \alpha_H) \Phi- (1+\alpha_T) \Psi + (\alpha_M - \alpha_T) H\pi  - \alpha_H \dot \pi = -\frac{\sigma_{\rm m}}{M^2}  \;,\label{Ein3}
\ee
while the trace of the same components gives,
using the  equation above,
\be
\begin{split}
&2 \ddot\Psi + 2 (3 + \alpha_M) H \dot  \Psi+ 2 (1 + \alpha_B ) H \dot \Phi   \\
& + 2 \left[   \dot H  - \frac{\rho_{\rm m} + p_{\rm m}}{2 M^2}  +  ( \alpha_B H)^{\hbox{$\cdot$}}+  (3 + \alpha_M)(1 + \alpha_B) H^2  \right] \Phi \\
&-2H\alpha_B\, \ddot \pi+ 2 \left[    \dot H + \frac{\rho_{\rm m} + p_{\rm m}}{ 2 M^2}  -  ( \alpha_B H)^{\hbox{$\cdot$}} - (3 + \alpha_M) \alpha_B H^2  \right] \dot \pi \\
& +2 \left[  (3+\alpha _M )H \dot H+\frac{\dot p_{\rm m}}{2 M^2}+ \ddot H \right]\pi = \frac{1}{M^2 } \left(\delta p_{\rm m} - \frac{2}{3} \frac{k^2}{a^2} \sigma_{\rm m} \right)\;.\label{Ein4}
\end{split}
\ee
The evolution equation for $\pi$ reads
\be
\begin{split}
&H^2\alpha_K \ddot \pi +\left\{ \left[ H^2(3+\alpha_M)+\dot H \right]\alpha_K+(H\alpha_K)^{\hbox{$\cdot$}} \right\} H\dot \pi \\
&+6\left\{ \left(\dot H+\frac{\rho_{\rm m}+ p_{\rm m}}{2M^2}\right)\dot H +\dot H \alpha_B \left[ H^2(3+\alpha_M)+\dot H \right]+H(\dot H \alpha_B)^{\hbox{$\cdot$}}\right\} \pi \\
&-2\frac{k^2}{a^2}\left\{ \dot H+\frac{\rho_{\rm m}+ p_{\rm m}}{2M^2}+H^2 \left[ 1+\alpha_B(1+\alpha_M)+\alpha_T-(1+\alpha_H)(1+\alpha_M) \right]+(H(\alpha_B-\alpha_H))^{\hbox{$\cdot$}}\right\} \pi \\
& + 6H\alpha_B \ddot \Psi +H^2(6\alpha_B-\alpha_K)\dot \Phi +6\left[\dot H+\frac{\rho_{\rm m}+ p_{\rm m}}{2M^2}+H^2\alpha_B(3+\alpha_M)+(\alpha_B H)^{\hbox{$\cdot$}}\right] \dot \Psi \\
&+\left[6\left(\dot H+\frac{\rho_{\rm m}+p_{\rm m}}{2M^2}\right)+H^2(6\alpha_B-\alpha_K)(3+\alpha_M)+2(9\alpha_B-\alpha_K)\dot H+H(6\dot\alpha_B-\dot \alpha_K)\right]H\Phi \\ 
&+2\frac{k^2}{a^2} \left\{ \alpha_H\dot \Psi + \left[H(\alpha_M+\alpha_H(1+\alpha_M)-\alpha_T)-\dot\alpha_H\right]\Psi+ (\alpha_H-\alpha_B)H\Phi  \right\}
 =0 \;. \label{sfee}
\end{split}
\ee
These equations have been previously derived in \cite{GLPV1}  in terms  of the effective field theory parameters.
Restricting to Horndeski theories ($\alpha_H=0$), they have  been also obtained in \cite{Bloomfield:2013efa} and later reproduced in \cite{Bellini:2014fua}, where the 
notation
used here was introduced. Note that as a consequence of the parameterization where $M^2$ is in factor of the full gravitational Lagrangian \eqref{lag-quad}, matter quantities always appear divided by a $M^2$ factor. In Appendix \ref{sec:lsl} we discuss the long wavelength behaviour of these equations for adiabatic initial conditions; in Appendix \ref{ref:app_syn} we write these 
 equations
 in synchronous gauge and conformal time, which is the coordinate system usually employed in CMB codes.

\subsection{Fluid description}
\label{sec:fluid}

It is sometimes convenient to describe the dark energy, both in the background and perturbative equations,  as an effective fluid.
In order to do so, we define the background energy density and pressure for dark energy, respectively, as
\be
\rho_D \equiv 3 M^2 H^2 - \rho_{\rm m} \;, \qquad p_D \equiv -M^2 (2 \dot H + 3 H^2) - p_{\rm m} \;.
\ee
These are simply derived quantities that can be computed once the evolution of the expansion history, the matter content and  the effective Planck mass $M$ are known.
With these definitions,  and using the conservation of the background matter stress-energy tensor,
\be
\dot \rho_{\rm m} + 3 H (\rho_{\rm m}+p_{\rm m}) =0\;, 
\ee
the conservation of the background dark energy stress-energy tensor reads
\be
\dot \rho_D = - 3 H (\rho_D+p_D) +3 \alpha_M M^2 H^3 =   3 H (\rho_{\rm m}+p_{\rm m}) + 6 M^2 H (\dot H + \alpha_M H^2 ) \;. 
\ee
Another useful relation that one can use to express $\dot p_D$ in terms of matter and geometry is
\be
\dot p_D = - \dot p_{\rm m} -  M^2 \big[ 2 \ddot H + 2 H \dot H (3 + \alpha_M)  + 3 \alpha_M  H^3  \big] \;,
\ee
which can be derived from the equations above.

Equations~\eqref{Ein1}--\eqref{Ein4} can then be rewritten in the usual form,
\begin{align}
       \frac{k^2 }{a^2} \Psi +3 H  \big( \dot \Psi + H \Phi \big)     &  =  - \frac{1}{ 2 M^2  } \sum_I \delta \rho _I \;,  \label{EE1fluid} \\
 \dot \Psi + H \Phi  &   =  - \frac{1}{ 2 M^2}\sum_I q_I  \;, \label{EE2fluid} \\
\Psi- \Phi &  = \frac{1}{  M^2 } \sum_I \sigma_I \;, \label{EE3fluid} \\
     \ddot \Psi + H \dot \Phi +  2\dot H \Phi          + 3H \big( \dot \Psi + H \Phi  \big)
& = \frac{1}{2 M^2 }  \sum_I \left( \delta p_I  -\frac23 \frac{k^2}{a^2} \sigma_I \right) \;,\label{EE4fluid}
\end{align}
where the sum is over the matter and the dark energy components. These equations implicitely define the quantities $\delta \rho_D$, $q_D$, $\delta p_D$ and $\sigma_D$ as the energy density perturbation, momentum, pressure perturbation and anisotropic stress of the dark energy fluid. An explicit definition is given 
in Newtonian gauge in Appendix \ref{sec:lsl}
and in synchronous gauge in Appendix 
\ref{ref:app_syn}.

 With these definitions,
one can verify that the evolution equation for $\pi$, eq.~\eqref{sfee},  
is equivalent to a conservation equation of the dark energy fluid quantities,
\be
\delta \dot \rho_D + 3 H (\delta \rho_D + \delta p_D) - 3 (\rho_D + p_D) \dot \Psi - \frac{k^2}{a^2} q_D =\alpha_M  H \sum_I \delta \rho_I \;.
\ee
The Euler equation,
\be
\dot q_D  +3 H  q_D + (\rho_D + p_D) \Phi +   \delta p_D - \frac23 \frac{k^2}{a^2} \sigma_D  = \alpha_M  H \sum_I  q _I\;,
\ee
is identically satisfied by the definitions of $q_D$, $\delta p_D$ and $\sigma_D$. 
Conservation of matter in the Jordan frame implies a continuity and Euler equations for matter with vanishing right-hand side.

To close the system, 
one needs to provide an equation of state for  dark energy or, at least, a relation between 
 $\delta p_D$ and $\sigma_D$ in terms of  $\delta \rho_D$, $q_D$ and the other matter variables.  
 In order to do so in the simpler case where $\alpha_H=0$,
 we solve eqs.~\eqref{Ein1}--\eqref{Ein3} for $\Psi$, $\dot \Psi$ and $\dot \pi$ and 
 then we
 plug these solutions in eqs.~\eqref{EE1fluid} and \eqref{EE2fluid} to express $\pi$ and $\Phi$ in terms of 
 $\delta \rho_{\rm m}$,  
 $q_{\rm m}$, $\sigma_{\rm m}$, $\delta \rho_D$ and $q_D$.  
 $\dot \Phi$ is obtained from the first derivative of \eqref{Ein3}. 
 To obtain $\ddot \Psi$ and $\ddot \pi$ we use eqs.~\eqref{Ein4} and \eqref{sfee}. Combining all these solutions we can finally  express $\sigma_D$ and $\delta p_D$ in terms of the other fluid variables. We obtain
\begin{align}
\delta p_D  = & \ \frac{\gamma_1 \gamma_2 + \gamma_3  \alpha_B^2 \tilde k^2}{\gamma_1 +  \alpha_B^2 \tilde k^2}  (\delta \rho_D-3H q_D)+\frac{\gamma_1 \gamma_4 + \gamma_5  \alpha_B^2 \tilde k^2}{\gamma_1 +  \alpha_B^2 \tilde k^2} H q_D \nonumber \\
& + \gamma_7 ( \delta \rho_{\rm m}  -3 H q_{\rm m}) +\frac{\gamma_1 \gamma_6 + 3 \gamma_7 \alpha_B^2 \tilde k^2 }{\gamma_1 +  \alpha_B^2 \tilde k^2} H q_{\rm m}  -\frac{6\alpha_B^2}\DD \delta p_{\rm m} \, ,\label{deltapd} \\
\sigma_D  =& \ \frac{a^2}{2 k^2}  \Bigg[ \frac{\gamma_1 \alpha_T + \gamma_8  \alpha_B^2 \tilde k^2}{\gamma_1 +  \alpha_B^2 \tilde k^2} (\delta \rho_D-3H q_D)+\frac{\gamma_9  \tilde k^2}{\gamma_1 +  \alpha_B^2 \tilde k^2} H q_D  \nonumber \\ 
&+ \alpha_T (\delta\rho_{\rm m}-3H q_{\rm m})+\frac{\gamma_{10}   \tilde k^2}{\gamma_1 +  \alpha_B^2 \tilde k^2} H q_{\rm m} \Bigg] \, ,\label{sigmad}
\end{align}
where we use the notation $\tilde k\equiv k/(aH)$ and
we have introduced dimensionless coefficients $\gamma_a$, whose expressions are explicitly given in Appendix \ref{sec:app_parameters}.
These relations for $\delta p_D$ and $\sigma_D$ are derived, to our knowledge,  for the first time and represent the most general description of dark energy in the context of Horndeski theories. In particular, eqs.~\eqref{deltapd} and \eqref{sigmad} extend the equations of state for perturbations derived in Refs.~\cite{Battye:2013aaa,Bloomfield:2013cyf}.

One can check that for adiabatic initial conditions, i.e.
\be
\pi \approx - (\dot \Psi + H \Phi)/\dot H \;, \qquad   \delta \rho_{\rm m} \approx \dot \rho_{\rm m} \pi\;, \qquad \delta p_{\rm m} \approx \dot p_{\rm m} \pi \;, \qquad v_{\rm m} \approx - \pi   \;,\label{adia1}
\ee
where the symbol $\approx$ denotes equality in the long wavelength limit $\tilde k \ll 1$, the dark energy equation of state satisfies
\be
\begin{split}
\delta \rho_D \approx - 3 H (\rho_D + p_D) \pi &\qquad  \delta p_D \approx - \dot p_D  \pi\;, \\
q_D \approx - (\rho_D + p_D) \pi \; ,& \qquad \sigma_D \approx - \alpha_T M^2 \Psi - (\alpha_T - \alpha_M) M^2 H \pi \;, \label{adia2}
\end{split}
\ee
which is what expected from the equations of motion  in Sec.~\ref{sec:Einstein}, see discussion in Appendix \ref{sec:lsl}.

Going back to arbitrary scales, let us discuss two illustrative examples.
\begin{itemize}
\item{ $\alpha_B =0$:} there is no braiding and the kinetic term of scalar fluctuations depends on $\alpha_K$ only,
$\alpha = \alpha_K$.
In this case eqs.~\eqref{deltapd} and \eqref{sigmad} reduce to 
\begin{align}
\delta p_D = & \ \tilde c_s^2 ( \delta \rho_D  -3 H q_D)-  \bigg[  \frac{\dot p_D + 3\alpha_M  H^3 M^2 }{\rho_D+p_D}  + H (\alpha_T- \alpha_M) \left( 1 - \frac{2 M^2}{3 (\rho_D + p_D)} \frac{k^2}{a^2} \right) \bigg]  q_D \nonumber \\
& + \frac{\alpha_T}{3} \delta \rho_{\rm tot} - \frac{\alpha_K}{6} (\alpha_T -\alpha_M) H q_{\rm m} \;, \\
 \sigma_D
= &  - \alpha_T  M^2 \Psi +  H (\alpha_T- \alpha_M) \frac{ M^2}{  \rho_D + p_D }  q_D \;,
\end{align}
where we have used 
eqs.~\eqref{EE1fluid},  \eqref{EE2fluid} and
$\tilde c_s^2 \equiv c_s^2 -2 (\alpha_T -\alpha_M)/\alpha_K = (\rho_D + p_D) /(\alpha_K  H^2 M^2 )$, $\delta \rho_{\rm tot} \equiv \delta \rho_{\rm m} + \delta \rho_D$. For $\alpha_T = 0 = \alpha_M$ we recover the standard $k$-essence pressure perturbation \cite{Garriga:1999vw} and 
no
anisotropic stress.  For $\alpha_T \neq 0$ or $\alpha_T - \alpha_M \neq 0$, the dark energy anisotropic stress is nonzero and simply given in terms of the total curvature $\Psi$ and 
the dark energy momentum $q_D$. 
Note that the term containing $k^2$ in the pressure perturbation  $\delta p_D$ cancels from the combination $\delta p_D - (2k^2/3a^2) \sigma_D$, which  appears as a source in the Euler equation and in the evolution equation for $\Psi$.

\item{ $\alpha_B^2 \gg \alpha_K$:} braiding dominates the time kinetic term, 
$\alpha \simeq 6 \alpha_B^2$.
However, 
one needs $\alpha_B \lesssim 1$ to avoid  gradient instabilities \cite{Creminelli:2008wc}.
In this case, from the definition of $\gamma_1$, eq.~\eqref{gamma1}, we have $\gamma_1 \simeq -3\alpha_B^2 \dot H/H^2$ so that, if we concentrate on  sub-horizon scales, $k\gg a H$, eqs.~\eqref{deltapd} and \eqref{sigmad} reduce to
\begin{align}
\delta p_D = &\left(c_s^2+\frac{\alpha_T}3+\frac{\xi}3-\frac{2\dot H+\ddot H/H-\xi \dot H}{\alpha_B } \frac{a^2}{k^2}\right)(\delta \rho_{D} - 3 H q_{D}) -\left(1+\xi \right)H q_D\nonumber \\
& -3\frac{\dot Ha^2}{k^2}\xi H q_{\rm m} -\left(1+\xi\right)\frac{\delta \rho_{\rm m} }3-\delta p_{\rm m}\;, \\
 \sigma_D
= & \ \xi \frac{a^2}{k^2} \left[ \frac{1}{2}(\delta \rho_{D} - 3 H q_{D}) + \frac32 H\alpha_B  q_{\rm tot}  \right] \;,
\end{align}
where 
$\xi\equiv ({\alpha_T-\alpha_M})/{\alpha_B}$ and $q_{\rm tot} \equiv q_{\rm m} + q_D$. As expected in this case \cite{Sawicki:2012re}, the behavior of dark energy is very different from that of a perfect fluid.
In particular,
 for $\tilde k^2 \lesssim \gamma_2/\gamma_3$ the relation between pressure and density perturbations 
is scale dependent.
For $\xi \neq 0$, 
the anisotropic stress is nonzero and has a 
scale dependence that differs from
the $\alpha_B=0$ case discussed above.
\end{itemize}

\subsection{Interface with the observations}
\label{sec:obs}

In Sec.~\ref{sec:Einstein} we have described the full set of evolution equations including the standard matter species directly using the scalar fluctuation $\pi$. 
These equations can be solved  in a modified Boltzmann code; for instance, 
they have been recently implemented in a code in \cite{Hu:2013twa,Hu:2014oga}. 
Alternatively, in Sec.~\ref{sec:fluid}
we have rewritten these equations in terms of dark energy fluid quantities and we have provided the full equations of state for Horndeski theories ($\alpha_H=0$), eqs.~\eqref{deltapd} and \eqref{sigmad}. In this approach, the equations of state fully encode the description of dark energy.
 
To discuss more easily the relation with late time observations we can 
use the Einstein equations in the fluid form, eqs.~\eqref{EE1fluid}--\eqref{EE4fluid}, and rewrite these two equations as an evolution equation for the gravitational potential $\Psi$ and a relation between $\Psi$ and $\Phi$. For simplicity,  we restrict again our discussion   to the case $\alpha_H=0$.  To do that we can first combine eqs.~\eqref{EE1fluid}--\eqref{EE3fluid} to solve for $\Phi$, $\delta \rho_D$ and $q_D$  in terms of $\Psi$, $\dot \Psi$, $\sigma_D$ and the matter field. Moreover, we can use eq.~\eqref{EE4fluid} to express $\delta p_D$ as a function of the other quantities.  Using these relations, it is straighforward to show that eqs.~\eqref{deltapd} and \eqref{sigmad} are equivalent to a dynamical equation for the gravitational potential $\Psi$, sourced by the matter fields,\footnote{An alternative derivation of eq.~\eqref{Phi_evol} is  to combine eqs.~\eqref{Ein1}--\eqref{Ein3} to solve for $\pi$, $\dot \pi$ and $\Phi$ in terms of $\Psi$, $\dot \Psi$ and the matter field. We can then use the time derivative of \eqref{Ein3} to solve for $\dot \Phi$ and the scalar field equation \eqref{sfee} to solve for $\ddot \pi$. Using these solutions, it is possible to eliminate the scalar field fluctuations and derive \eqref{Phi_evol} \cite{Bellini:2014fua}. We can then derive eq.~\eqref{Phitopsi} from \eqref{Ein3}.}
\beq
\begin{split}
&\ddot \Psi +  \frac{\mbeta_1 \nbeta_2 + \nbeta_3  \alpha_B^2 \,\tilde k^2}{\mbeta_1+\alpha_B^2 \tilde k^2} H \dot \Psi+   \frac{ \mbeta_1 \nbeta_4+\mbeta_1 \nbeta_5  \,\tilde k^2   +c_s^2  \alpha_B^2 \tilde k^4} {\mbeta_1+ \alpha_B^2 \,\tilde k^2 }  H^2  \Psi = - \frac{1}{2M^2} \Bigg[  \frac{\mbeta_1\nbeta_6 + \nbeta_7  \alpha_B^2 \, \tilde k^2  }{\mbeta_1+ \alpha_B^2 \,\tilde k^2  }  \delta \rho_{\rm m}  \\
&+ \frac{\mbeta_1 \nbeta_8 + \nbeta_9  \alpha_B^2\, \tilde k^2  }{\mbeta_1+ \alpha_B^2 \,\tilde k^2 } H q_{\rm m} + \frac{ \mbeta_1 \nbeta_{10} + \mbeta_1 \nbeta_{11}\,   \tilde k^2   +\frac23  \alpha_B^2 \tilde k^4 }{\mbeta_1+ \alpha_B^2\, \tilde k^2  } H^2  \sigma_{\rm m}   - \frac{\alpha_K}{\DD} \delta p_{\rm m}- 2 H \dot \sigma_{\rm m} \Bigg]\;, \label{Phi_evol}
\end{split}
\eeq
where the  dimensionless parameters $\nbeta_a$ are explicitly given in Appendix \ref{sec:app_parameters},
and a relation between $\Phi$ and $\Psi$, involving $\dot \Psi$ and the matter fields,
\beq
\label{Phitopsi}
\begin{split}
&\alpha_B^2 \tilde k^2 \left[ \Phi - \Psi \left( 1 + \alpha_T + \frac{ 2 \gamma_9  }{ \DD \alpha_B} \right) + \frac{\sigma_{\rm m}}{M^2} \right]  + \mbeta_1  \left[ \Phi - \Psi ( 1 + \alpha_T ) \frac{\gamma_1}{\beta_1}
+ \frac{\sigma_{\rm m}}{M^2} \right] = \\
&\frac{\gamma_9}{ H^2  M^2}    \left[ \frac{\alpha_B}{\DD} \,  ( \delta \rho_{\rm m} -3 H q_{\rm m} )+  H M^2  \, \dot \Psi + H \frac{\alpha_K }{2 \DD} \, q_{\rm m}  -  H^2  \, \sigma_{\rm m}   \right] \;.
\end{split}
\eeq
Combined with the evolution equations for matter, these  equations  form a close system.
They generalize those given in  \cite{Bellini:2014fua}, which we recover for  $\delta p_{\rm m} = 0 =\sigma_{\rm m} $.
Following \cite{Bellini:2014fua},  the parameter $\beta_1$
appears in eq.~\eqref{Phi_evol} to make explicit 
the existence of a transition scale in the dynamics, $k_B\equiv a H \mbeta_1^{1/2}/ \alpha_B$, which  has been called {\em braiding scale}. Here we find that for $\alpha_T \neq \alpha_M$ this scale is different from the transition scale $a H \gamma_1^{1/2}/ \alpha_B$ appearing in eqs.~\eqref{deltapd} and \eqref{sigmad}. In particular, $\mbeta_1$
is related to  $ \gamma_1$ by
 \be
\mbeta_1 = \gamma_1 - \gamma_9 \;, 
\ee  
(see the explicit definition in terms of the $\alpha_i$ in eq.~\eqref{beta1}). Note that eq.~\eqref{Phitopsi} displays both scales.

Let us consider again the two limits discussed before (see also \cite{Bellini:2014fua}).
\begin{itemize}

\item{ $\alpha_B =0$:}
In this case most of the scale dependences go away. We are left with the simpler expression
\beq
\begin{split}
&\ddot \Psi +  \left( 4+ 2 \alpha_M + 3 \Upsilon\right)H \dot \Psi+  \left(\nbeta_4H^2+c_s^2 \frac{k^2}{a^2}  \right)  \Psi =\\
& - \frac{1}{2M^2} \Bigg[c_s^2  (\delta \rho_{\rm m}-3 H q_{\rm m})  + ( \alpha_M  - \alpha_T+  3 \Upsilon)  H q_{\rm m} + \left( \nbeta_{10} H^2+ \frac{2k^2}{3a^2}\right)   \sigma_{\rm m}   -  \delta p_{\rm m}+ 2 H \dot \sigma_{\rm m} \Bigg]\;. \label{Phi_evol_nob}
\end{split}
\eeq
Although both $\alpha_M$ and $\alpha_T$ can be nonzero here, the form of this equation is very similar to that obtained in the standard $k$-essence case.

\item{ $\alpha_B^2\gg \alpha_K$:}
For simplicity we consider only the case $\alpha_T=0$. Moreover, to avoid negative gradient instabilities we require $\alpha_B \lesssim {\cal O}(1)$ \cite{Creminelli:2008wc}. However, no such a restriction is imposed on  $\alpha_M$, whose value can 
affect 
the braiding scale. Indeed, when $\alpha_B^2\gg \alpha_K$, this is given by
\be
\frac{k_B^2}{a^2}\simeq 3(H^2\alpha_M-\dot H)\; .
\ee
Considering modes with $k\gg k_B$, eq.~\eqref{Phi_evol} simplifies to
\beq
\begin{split}
&\ddot \Psi +  (3+\alpha_M) H \dot \Psi+ \left( \frac{k_B^2 \beta_5}{a^2}  + c_s^2 \frac{k^2}{a^2} \right)\Psi \simeq  - \frac{1}{2M^2} \left(\frac{ k_B^2 \beta_6 }{ k^2}+ c_s^2+\frac13-\frac{\alpha_M}{3\alpha_B}\right)  \delta \rho_{\rm m}  \;, \label{Phi_evol_b}
\end{split}
\eeq
where we have neglected relativistic terms on the right hand side of \eqref{Phi_evol}. The mass scale ${k_B^2 \beta_5}/{a^2}$ corresponds to the so-called Compton mass. Depending on the value of $\beta_5$, this scale may be inside the horizon and induce a transition on the behaviour of the effective Newton constant, which is considered a strong signal of modified gravity.

\end{itemize}

To make the link with observations without resorting to numerical calculations, one often relies on the quasi static approximation, which corresponds to neglecting time derivatives  with respect to spatial ones  on scales much below the sound horizon, i.e.~for $k\gg aH/c_s $. In this regime, modifications of gravity can be captured by two quantities, the effective Newton constant $G_{\rm eff}$, defined by
\be
-\frac{k^2}{a^2}\Phi=4\pi G_{\rm eff} \delta \rho_{\rm m}\; ,
\ee
and the gravitational slip $\gamma \equiv \Psi/\Phi$. 

Both these quantities can be computed using eqs.~\eqref{Phi_evol} and~\eqref{Phitopsi}. However, as discussed in \cite{Bellini:2014fua} this does not give the same result 
as neglecting time derivatives in  eq.~\eqref{sfee} and using eqs.~\eqref{Ein1} and \eqref{Ein3} to derive $G_{\rm eff}$ and $\gamma$.
The two procedures
are  consistent if $k$ is  much larger than the other scales, 
i.e.~in the limit $k\rightarrow \infty$. In this case we recover (compare for instance with the results of  \cite{GLPV1} in the same limit)
\be
8\pi G_{{\rm eff}}=\frac{\DD\,  c_s^2(1+\alpha_T) +2 \left[ \alpha_B (1+\alpha_T) + \alpha_T - \alpha_M \right]^2}{\DD\,  c_s^2}\, M^{-2}\; , 
\ee
\be
\gamma= \frac{\DD\,  c_s^2 +2\alpha_B \left[\alpha_B (1+\alpha_T) +\alpha_T -\alpha_M \right]}{\DD\,  c_s^2(1+\alpha_T) +2 \left[ \alpha_B (1+\alpha_T) + \alpha_T - \alpha_M \right]^2}\;, 
\ee
where we have expressed both quantities directly in terms of the functions $\alpha_i$ (recall that 
$\DD=\alpha_K+6\alpha_B^2$ and $\alpha_H$ is here set to zero), obtained from the derivatives of the initial ADM Lagrangian.

\section{Conclusions}
In this article, we have presented a very general approach to parametrize theoretically motivated deviations from the $\Lambda$CDM standard model. This approach combines several advantages, both from the  theoretical and observational points of view. On the one hand, it provides a unified treatment of theoretical models,  using as a starting point a Lagrangian expressed in terms of ADM geometrical quantities defined  for a foliation of uniform scalar field hypersurfaces. On the other hand, it expresses all the relevant  information about the linear cosmological  perturbations in terms of a 
minimal set of
five time-dependent functions, which can be constrained by observations. These five functions, together with the background time evolution (and a constant parameter), are sufficient to fully characterize the background and linear perturbations, within the large class of models we have considered (corresponding to the conditions (\ref{conds}) to avoid a non trivial dispersion relation for the scalar mode). 

The link between these two endpoints, theoretical and observational, is direct since the five functions correspond to combinations of the derivatives of the initial Lagrangian. One can thus automatically derive the observational predictions for any existing or novel model by computing these functions from the ADM Lagrangian. Conversely, one can use this approach in a model-independent way by trying to constrain the five arbitrary functions (this requires  some parametrization of these free functions expressed for instance in terms of the redshift;
see e.g.~\cite{Piazza:2013pua})
with observations. Of course, since  the bounds on parameters are less stringent as the number of parameters increases, it would be interesting to analyse the data with a scale of  increasing complexity, corresponding to the number of free functions,  thus covering the range  from the simplest theory, i.e. $\Lambda$CDM, where all functions are zero,  to  more  and more general theories.

\vspace{0.5cm}

{\bf Acknowledgements:} 
We are particularly 
indebted to 
 Federico Piazza for uncountable inspiring discussions and 
initial collaboration on this project. We would also like to thank Michele Mancarella and Enrico Pajer for useful discussions. D.L. is partly supported by the ANR (Agence Nationale de la Recherche) grant STR-COSMO ANR-09-BLAN-0157-01. J.G.~and F.V.~acknowledge financial support from {\em Programme National de Cosmologie et Galaxies} (PNCG) of CNRS/INSU, France and thank PCCP and APC for kind hospitality. 

\appendix

\section{Superhorizon evolution}
\label{sec:lsl}
In this appendix we extend the arguments of  \cite{Weinberg:2003sw,Bertschinger:2006aw}  and check  that the Einstein equations of Sec.~\ref{sec:Einstein} 
satisfy the usual adiabatic solution on superhorizon scales.
To this end, it is convenient to define the quantities 
\be
{\cal P} \equiv M^2( \dot \pi - \Phi ) \;, \qquad {\cal Q} \equiv M^2( \dot \Psi + H \Phi + \dot H \pi ) \;, \qquad {\cal R} \equiv M^2(  \Psi + H \pi) \;,
\ee
(note that $Q = \dot {\cal R} - H ( {\cal P} + \alpha_M {\cal R})$) and rewrite the Einstein equations \eqref{Ein1}--\eqref{Ein4} respectively as
\begin{align}
-2 \frac{k^2}{a^2} \left[ (1+ \alpha_H) {\cal R} - (1+ \alpha_B) M^2 H  \pi  \right]  & \nonumber \\  -  6 H (1+\alpha_B) {\cal Q} - H^2 (\alpha_K - 6 \alpha_B) {\cal P} & =  \delta \rho_{\rm m} - \dot \rho_{\rm m} \pi \;,
\label{EE1} \\
- 2 {\cal Q} + 2 \alpha_B H {\cal P} & =   (\rho_{\rm m} + p_{\rm m}) (v_{\rm m} + \pi) \;,
\label{EE2} \\
M^2( \Psi - \Phi ) + \alpha_T {\cal R} - \alpha_M H \pi + \alpha_H  {\cal P} & =  \sigma_{\rm m} \;,
\label{EE3} \\
2 \dot  {\cal Q} +  6 H  {\cal Q} + 2 \bigg(  \frac{\rho_{\rm m} + p_{\rm m}}{2 M^2} -3 \alpha_B H^2  \bigg) {\cal P} - 2  (H \alpha_B {\cal P})^{\hbox{$\cdot$}} &=   \delta p_{\rm m} - \dot p_{\rm m} \pi - \frac{2}{3} \frac{k^2}{a^2} \sigma_{\rm m}  \;.
\label{EE4}
\end{align}
The evolution of $\pi$ reads
\be
\begin{split}
&(H^2 \alpha_K {\cal P})^{\hbox{$\cdot$}} + 6 ( H \alpha_B {\cal Q} )^{\hbox{$\cdot$}}  + 3 H  ( \alpha_K H^2 -2 \alpha_B \dot H) {\cal P} + 36\bigg(   \dot H + \frac{\rho_{\rm m} + p_{\rm m}}{2 M^2} +3  H^2 \alpha_B \bigg)  {\cal Q}  \\
&+ \frac{k^2}{a^2} \bigg\{ 2  \alpha_H {\cal Q} + 2 \Big[  H \alpha_H + (M^2 \alpha_H)^{\hbox{$\cdot$}} M^{-2} +  H \alpha_M -  H \alpha_T  \Big]{\cal R} \\
& - 2 \bigg[  \dot H + \frac{ \rho_{\rm m} + p_{\rm m}}{2 M^2} +  H^2 \alpha_B +  (M^2 H \alpha_B)^{\hbox{$\cdot$}}M^{-2} \bigg] M^2 \pi -2 M^2 H \alpha_B \Phi \bigg\} =0 \;.
\end{split}
\label{pi_evol}
\ee
Moreover, in terms of these quantites, the definitions of the fluid variables introduced in eqs.~\eqref{EE1fluid}--\eqref{EE4fluid} are given by
\begin{align}
\delta \rho_D &\equiv 2  \frac{k^2}{a^2} \big( \alpha_H {\cal R} - \alpha_B M^2 H \pi \big) - {3 H \big[ (\rho_D + p_D) \pi -2  \alpha_B {\cal Q} \big]} +  H^2 (\alpha_K - 6 \alpha_B) {\cal P} \;, \label{dec1} \\
q_D & \equiv   - { 2 \alpha_B  H {\cal P} }- (\rho_D + p_D) \pi \;, \label{dec2} \\
\sigma_D & \equiv {  \alpha_M M^2 H \pi - \alpha_T {\cal R} - \alpha_H {\cal P} } \;, \label{dec3} \\
\delta p_D & \equiv  \big[ \dot p_D +  \alpha_M H M^2 (2 \dot H + 3 H^2 ) \big] \pi - 2 \alpha_M  H {\cal Q} \nonumber \\
& + \bigg( \frac{\rho_D + p_D}{M^2} + 6 \alpha_B H^2   \bigg) {\cal P} + 2 \big( \alpha_B  H {\cal P} \big)^{\hbox{$\cdot$}} + \frac23 \frac{k^2}{a^2} \sigma_D \;.\label{dec4}
\end{align}

Independently of the constituents of the Universe, the $k=0$ mode of the field equations for scalar fluctuations in Newtonian gauge is invariant under the coordinate transformation \cite{Weinberg:2003sw}
\begin{align}
t &\to t + \epsilon (t) \;, \\
x^i &\to x^i (1-\lambda) \;, \label{space_coord}
\end{align} 
where $\epsilon$ is an arbitrary infinitesimal function of time and $\lambda$ an arbitrary infinitesimal constant. In particular, using these transformations one can start from an unperturbed FLRW solution and generate a solution in Newtonian gauge with metric perturbations
\be
\Psi = H \epsilon -\lambda \;, \qquad \Phi = -\dot \epsilon \;, \label{Psi_Phi_ls}
\ee
and, assuming the Universe filled by several fluids and scalar fields, with matter perturbations
\be
\delta \rho_X = - \dot \rho_X \epsilon\;, \qquad \delta \varphi_X = - \dot \varphi_X \epsilon\;. \label{adia_fluctuations}
\ee
Note that these solutions remain valid also if individual matter components are not separately conserved \cite{Weinberg:2003sw}, as  in the case of dark energy components that are non-minimally coupled to gravity.

Let us check that this is solution of the above equations. For our dark energy and matter components, eq.~\eqref{adia_fluctuations} becomes
\be
\delta \rho_{\rm m} = - \dot \rho_{\rm m} \epsilon \;, \qquad \pi = - \epsilon \;. \label{adia_fluctuations2}
\ee
This second equality, together with eq.~\eqref{Psi_Phi_ls}, implies that ${\cal P} = 0$ and ${\cal Q} =0$ from which it follows that 
eqs.~\eqref{EE1}, \eqref{EE4} and the evolution equation for $\pi$, eq.~\eqref{pi_evol}, are satisfied  for $k=0$.

The remaining equations, i.e.~\eqref{EE2} and \eqref{EE3}, are  automatically satisfied because they multiply an overall factor of $k$ and $k^2$, respectively, that has been dropped here.
However, for the solutions \eqref{Psi_Phi_ls} and \eqref{adia_fluctuations} to be physical we must  require that these equations are satisfied for finite $k$ in the $k \to 0$ limit \cite{Weinberg:2003sw}, which implies
\be
v_{\rm m} = - \pi  \;, \label{adia}
\ee
and 
\be
( M^2  \epsilon )^{\hbox{$\cdot$}} +  H M^2   \epsilon  - \sigma_{\rm m} = M^2 (1+\alpha_T) \lambda \;,
\ee
with solution 
\be
\epsilon =  \frac{1}{M^2 a } \int^{t}   a \left[ M^2 (1+\alpha_T) \lambda + \sigma_{\rm m} \right] dt' \;. \label{eps_sol}
\ee

Equations \eqref{Psi_Phi_ls} and \eqref{adia_fluctuations2}, with the conditions \eqref{adia} and \eqref{eps_sol} correspond to the well-known super-horizon adiabatic solution. One can define the  quantity $\zeta$ from the metric perturbation as

\be
\zeta_{\rm tot} \equiv - \Psi + H \frac{\dot \Psi + H \Phi}{\dot H}\;,
\ee

which is known to be conserved in the $k \to 0$ limit for adiabatic perturbations \cite{Bardeen:1983qw}.
Indeed, one can replace  the solutions \eqref{Psi_Phi_ls} in its definition and check that in this limit it coincides with the constant $\lambda$ in eq.~\eqref{space_coord},
\be
\zeta_{\rm tot}= \lambda = - {\cal R} M^{-2}\;, \qquad k \to 0\;.
\ee
We note that on super-Hubble scales and for adiabatic initial conditions, eq.~\eqref{adia1}, eqs.~\eqref{Phi_evol}-\eqref{Phitopsi} 
reduce to the conservation of the total comoving curvature perturbation, i.e., $\dot \zeta_{\rm tot} \approx 0$.

\section{Evolution equations in synchronous gauge}
\label{ref:app_syn}

For completeness, in this section we give the perturbation equations in synchronous gauge. In this gauge, the perturbed FLRW metric including scalar perturbations 
reads
\be
ds^2=-dt^2+a^2(t)(\delta_{ij}+h_{ij})dx^idx^j\, ,
\ee
with
\be
h_{ij} \equiv \frac{1}3h \delta_{ij}+\left( \frac{ k_i k_j}{k^2} -\frac{1}3\delta_{ij} \right) \left(h+6\eta\right)\, .
\ee
Using  gauge transformations (see for instance \cite{Ma:1995ey}), we can express the variables in Newtonian gauge into those in synchronous gauge, which yields
\be
\begin{split}
\pi^{(N)}&=\pi^{(S)}+ \delta t\, , \\ 
\Phi &= \delta \dot t \, , \\ 
\Psi &=\eta-H\delta t\, ,\\
\quad \delta \rho_{\rm m}^{(N)}&= \delta \rho_{\rm m}^{(S)}-3H(\rho_{\rm m}+p_{\rm m})\delta t\, ,\\ 
\delta p_{\rm m}^{(N)} & = \delta p_{\rm m}^{(S)}+\dot p_{\rm m}\delta t \, , \\
\quad v_{\rm m}^{(N)} &=-\frac{\theta_{\rm m}^{(S)}}{k^2}-\delta t \, ,
\end{split}
\ee
with
\be
\delta t \equiv \frac{a^2}{k^2}\big(\dot h+6\dot \eta \big) \;,
\ee
and where we have introduced the divergence of the velocity, $\theta \equiv \vec \nabla \cdot \vec v $, related to the velocity potential, in Fourier space, by $\theta = - k^2 v$.
The anisotropic stress is gauge invariant.

We can now apply these gauge transformations to eqs.~\eqref{Ein1}--\eqref{Ein4}. We will use conformal time, $\tau \equiv \int dt/a$, which is usually employed in numerical codes.
Using this time, it is convenient to rescale the scalar fluctuation $\pi$ and the velocity divergence $\theta$ by the conformal factor and redefine 
\be
\pi \to {\pi}/a\, , \quad \theta \to {\theta}/a\, . 
\ee
By denoting by a prime the derivative with respect to conformal time, the Einstein equations in synchronous gauge read
(\noindent$(00)$ component)
\be
\begin{split}
&2k^2(1+\alpha_H)\eta-{\cal H}(1+\alpha_B)h'-{\cal H}^2(6\alpha_B-\alpha_K)\pi'\\
&+{\cal H}\left[2k^2(\alpha_H-\alpha_B)+{\cal H}^2(\alpha_K-12\alpha_B)+6{\cal H}'\alpha_B\right]\pi=-\frac{a^2}{M^2}\left[\delta\rho_{\rm m}-3{\cal H}(\rho_D+p_D)\pi\right]\, ,
\end{split}
\ee
($(0i)$ component)
\be
\begin{split}
\eta'-{\cal H}\alpha_B\pi'-{\cal H}^2\alpha_B\pi=\frac{a^2}{2M^2}\big[(\rho_{\rm m}+p_{\rm m}){\theta_{\rm m}}/{k^2}+(\rho_D+p_D)\pi \big]\, ,
\end{split}
\ee
($(ij)$-traceless)
\be
\begin{split}
h''+6\eta''+{\cal H}(2+\alpha_M)(h'+6\eta')-2k^2(1+\alpha_T)\eta-2k^2\alpha_H \pi' -2k^2{\cal H}\left(\alpha_H+\alpha_T-\alpha_M\right)=-\frac{2k^2}{M^2}\sigma_{\rm m}\, ,
\end{split}
\ee
and 
($(ii)$-trace)
\be
\begin{split}
&h''+{\cal H}(2+\alpha_M)h'-2k^2(1+\alpha_T)\eta +6\alpha_B\pi''+2\left[3{\cal H}^2\alpha_B(3+\alpha_M)+(3\alpha_B{\cal H})'-k^2\alpha_H\right]\pi' \\
&\left\{3{\cal H}^2[3\alpha_M+2\alpha_B(2+\alpha_M) ]+6\alpha_B {\cal H}'+6(\alpha_B {\cal H})'-2k^2(\alpha_H+\alpha_T-\alpha_M)\right\}{\cal H}\pi \\
&=\frac{a^2}{M^2}\left[-3(\rho_D+p_D)(\pi'+{\cal H}\pi)-3 p_D' \pi-3\delta p_{\rm m} -2\frac{k^2}{a^2} \sigma_{\rm m}\right]\, ,
\end{split}
\ee
where 
\be
\rho_D + p_D = - \rho_{\rm m} - p_{\rm m} -2 \frac{M^2}{a^2} \big( {\cal H} ' - {\cal H}^2 \big) \;,  \qquad {\cal H} \equiv \frac{a'}{a}\;.
\ee
The evolution equation for the scalar fluctuation reads
\be
\begin{split}
&-{\cal H}^2 \alpha_K \pi'' -\left[{\cal H}^2\alpha_K(2+\alpha_M)+{\cal {\cal H}}'\alpha_K+(\alpha_K{\cal H})'\right]{\cal H}\pi'\\
&+2k^2\left\{H^2\big[ (\alpha_B-\alpha_{\cal H})\alpha_M+\alpha_T-\alpha_M\big]+\left[(\alpha_B-\alpha_{\cal H}){\cal H}\right]'-\frac{a^2}{2M^2}(\rho_D+p_D)\right\}\pi\\&+\bigg\{{\cal H}^4[ 6\alpha_B\alpha_M-\alpha_K(1+\alpha_M) ]
-3{\cal H}^2{\cal H}' [\alpha_K-2\alpha_B(3-\alpha_M)]\\
&-6\alpha_B {\cal H}'^2+{\cal H}^3(6\alpha_B'-\alpha_K')-6{\cal H} ( \alpha_B{\cal H}' )' -\frac{3a^2}{2M^2}(\rho_D+p_D)({\cal H}^2-{\cal H}')\bigg\}\pi\\
&+{\cal H}\alpha_B h'' -2 k^2\alpha_H \eta' +\left[{\cal H}^2\alpha_B(1+\alpha_M)+(\alpha_B {\cal H})'-\frac{a^2}{2M^2}(\rho_D+p_D)\right]h'\\
&-2k^2\left\{ {\cal H} [\alpha_M+\alpha_H(1+\alpha_M)-\alpha_T]+\alpha_{\cal H}'\right\} \eta=0 \, .
\end{split}
\ee

Moreover, we can write these equations in terms of fluid quantities,
\begin{align}
  k^2\eta -\frac{1}2 {\cal H}h'    &  =  - \frac{a^2}{ 2 M^2  } \sum_I \delta \rho _I \;,  \label{EE1SC} \\
k^2 \eta'    & =   \frac{a^2}{ 2 M^2}\sum_I (\rho_I+p_I)\theta_I  \;, \label{EE2SC} \\
h''+6\eta'' +2{\cal H}(h'+6\eta')-2k^2\eta  & =- \frac{2 k^2}{  M^2 } \sum_I \sigma_I \;, \label{EE3SC} \\
   h''+2{\cal H} h'-2k^2\eta & =- \frac{3}{ M^2 }  \sum_I \left( \delta p_I  +\frac23 \frac{k^2}{a^2} \sigma_I \right) \;,\label{EE4SC}
\end{align}
where we have defined 
\begin{align}
\delta \rho_D&\equiv \frac{M^2}{a^2}\bigg\{2k^2\alpha_H\eta-{\cal H}\alpha_Bh' +{\cal H}^2 (\alpha_K-6\alpha_B)\pi'+\bigg[-2k^2(\alpha_B-\alpha_{\cal H})+6{\cal H}'\alpha_B \nonumber\\
&+{\cal H}^2(\alpha_K-12\alpha_B)-3\frac{a^2}{M^2}(\rho_D+p_D)\bigg]{\cal H}\pi\bigg\}\, ,\\
\theta_D&\equiv \frac{k^2M^2}{a^2(\rho_D+p_D)}\left\{ 2{\cal H}\alpha_B\pi'+\left[2{\cal H}^2+\frac{a^2}{M^2}(\rho_D+p_D)\right]\pi\right\} \, , \\
\sigma_D&\equiv-\frac{M^2}{k^2}\left[k^2\eta\alpha_T-\frac{{\cal H}\alpha_M}2\left(h'+6\eta'\right)+k^2\alpha_H\pi'+k^2\left(\alpha_H+\alpha_T-\alpha_M\right){\cal H}\pi\right]\, ,\\
\delta p_D&\equiv\frac23 \frac{k^2}{a^2} \sigma_D^S+\frac{M^2}{a^2}\bigg\{-2{\cal H}\alpha_M\eta'+2{\cal H}\alpha_B\pi''+\left[2[{\cal H}^2\alpha_B(3+\alpha_M)+(\alpha_B{\cal H})']+\frac{a^2}{M^2}(\rho_D+p_D)\right]\pi'\nonumber\\
&+\left[{\cal H}^3\left[3\alpha_M+2\alpha_B(2+\alpha_M)\right]+2{\cal H} {\cal H}'\alpha_B+2{\cal H}(\alpha_B {\cal H})'+\frac{a^2}{M^2}\left[{\cal H}(\rho_D+p_D)+p_D'\right]\right]\pi\bigg\}\, .
\end{align}
The equation for $\pi$ is equivalent to the continuity equation in conformal synchronous gauge, namely
\be
\delta \rho_D'+3{\cal H}(\delta \rho_D+\delta p_D)+(\rho_D+p_D) \left(\theta_D + \frac{h'}{2} \right)={\cal H}\alpha_M \sum_I \delta \rho _I\, ,
\ee
and the Euler equation
\be
\theta_D'+{\cal H}\left[1 + \frac{p_D'}{{\cal H} (\rho_D + p_D)} + \alpha_M \sum_I \frac{\rho_I}{\rho_D+p_D} \right] \theta_D+\frac{1}{\rho_D + p_D} \left (\delta p_D-\frac23 \frac{k^2}{a^2} \sigma_D \right) ={\cal H}\alpha_M \sum_I \frac{\rho_I+p_I}{\rho_D+p_D}\theta_I\, ,
\ee
 is an identity just as in the Newtonian gauge case.

\section{Scale dependence and definitions of the parameters}
\label{sec:app_parameters}

We report here the two ``equations of state'' for the dark energy fluid, eqs.~\eqref{deltapd} and \eqref{sigmad},
\begin{align}
\delta p_D  = & \ \frac{\gamma_1 \gamma_2 + \gamma_3  \alpha_B^2 \tilde k^2}{\gamma_1 +  \alpha_B^2 \tilde k^2}  (\delta \rho_D-3H q_D)+\frac{\gamma_1 \gamma_4 + \gamma_5  \alpha_B^2 \tilde k^2}{\gamma_1 +  \alpha_B^2 \tilde k^2} H q_D \nonumber \\
& + \gamma_7 ( \delta \rho_{\rm m}  -3 H q_{\rm m}) +\frac{\gamma_1 \gamma_6 + 3 \gamma_7 \alpha_B^2 \tilde k^2 }{\gamma_1 +  \alpha_B^2 \tilde k^2} H q_{\rm m}  -\frac{6\alpha_B^2}\DD \delta p_{\rm m} \, , \\
\sigma_D  =& \ \frac{a^2}{2 k^2}  \Bigg[ \frac{\gamma_1 \alpha_T + \gamma_8  \alpha_B^2 \tilde k^2}{\gamma_1 +  \alpha_B^2 \tilde k^2} (\delta \rho_D-3H q_D)+\frac{\gamma_9  \tilde k^2}{\gamma_1 +  \alpha_B^2 \tilde k^2} H q_D  \nonumber \\ 
&+ \alpha_T (\delta\rho_{\rm m}-3H q_{\rm m})+\frac{\gamma_{10}   \tilde k^2}{\gamma_1 +  \alpha_B^2 \tilde k^2} H q_{\rm m} \Bigg] \, ,
\end{align}
and 
provide the definitions of the parameters $\gamma_a$, 
for which we have assumed $\alpha_H=0$:
\begin{align}
\gamma_1 &\equiv \alpha_K \frac{\rho_D+p_D}{4 H^2 M^2 } -3 \alpha_B^2 \frac{\dot H}{H^2}  \;, \label{gamma1} \\
\gamma_2 & \equiv c_s^2 + \frac{\alpha_T}{3} -2 \frac{\alpha_B (2+ \Gamma)+ (1+\alpha_B) (\alpha_M - \alpha_T) }{\DD} \;, \\
\gamma_3 & \equiv c_s^2  + \frac{\gamma_8}{3 }\;, \\
\gamma_4 & \equiv \frac{1}{\rho_D+p_D} \bigg\{ -  \dot p_D/H   +\alpha_M \big[ \rho_D +p_D- 3  H^2 M^2 \big]  \nonumber  \\
& +6 \frac{\alpha_B^2}{\DD} \Big[(3 + \alpha_M + \Gamma ) (\rho_{\rm m}+p_{\rm m})-\dot p_{\rm m}/H   \Big] \bigg\}  \;, \\
\gamma_5 & \equiv -1-\frac{(6\alpha_B-\alpha_K)(\alpha_T- \alpha_M)}{6\alpha_B^2} +  \frac{\alpha_B^2}{ H\DD} \left(\frac{ \alpha_K}{\alpha_B^2}\right)^{\hbox{$\cdot$}} \\
\gamma_6 & \equiv  -  6 \alpha_B^2 \frac{2+ \Gamma }{\DD} +\frac{\alpha_K \alpha_M - 6 \alpha_B^2}{ \DD}  \;, \\
\gamma_7 & \equiv \frac{\alpha_K \alpha_M - 6 \alpha_B^2}{3 \DD}- \frac{(6\alpha_B-\alpha_K)(\alpha_T-\alpha_M)}{3\DD}  \;, \\
\gamma_8& \equiv\alpha_T +\frac{\alpha_T-\alpha_M}{\alpha_B}\; , \\
\gamma_9 & \equiv \DD\frac{\alpha_T-\alpha_M}{2 }\; , \\ 
\gamma_{10} & \equiv3  \alpha_B^2 (\alpha_T-\alpha_M)\; , 
\end{align}
where 
\be
\gamma_1  \Gamma \equiv\frac{\alpha_K}{4 H^2 M^2}\left[(3+\alpha_M)(\rho_{\rm m}+p_{\rm m})-\dot p_{\rm m}/H-\frac{\alpha_B^2 (\rho_D+p_D)}{\alpha_K H } \left(\frac{ \alpha_K}{\alpha_B^2}\right)^{\hbox{$\cdot$}} \right]- \DD\frac{\ddot H}{2H^3}\; ,
\ee
and
\beq
c_s^2 = - \frac{2(1+\alpha_B) \Big[ \dot H -  (\alpha_M - \alpha_T) H^2 + H^2 \alpha_B (1+\alpha_T)\Big] +2 H \dot \alpha_B + (\rho_{\rm m} + p_{\rm m})/M^2}{H^2 \DD}\;.
\eeq

As explained in Sec.~\ref{sec:obs},  the equations of state are equivalent to the dynamical equation for $\Psi$, eq.~\eqref{Phi_evol} and eq.~\eqref{Phitopsi},
\begin{align}
&\ddot \Psi +  \frac{\mbeta_1 \nbeta_2 + \nbeta_3  \alpha_B^2 \,\tilde k^2}{\mbeta_1+\alpha_B^2 \tilde k^2} H \dot \Psi+   \frac{ \mbeta_1 \nbeta_4+\mbeta_1 \nbeta_5  \,\tilde k^2   +c_s^2  \alpha_B^2 \tilde k^4} {\mbeta_1+ \alpha_B^2 \,\tilde k^2 }  H^2  \Psi = - \frac{1}{2M^2} \Bigg[  \frac{\mbeta_1\nbeta_6 + \nbeta_7  \alpha_B^2 \, \tilde k^2  }{\mbeta_1+ \alpha_B^2 \,\tilde k^2  }  \delta \rho_{\rm m}  \nonumber \\
&+ \frac{\mbeta_1 \nbeta_8 + \nbeta_9  \alpha_B^2\, \tilde k^2  }{\mbeta_1+ \alpha_B^2 \,\tilde k^2 } H q_{\rm m} + \frac{ \mbeta_1 \nbeta_{10} + \mbeta_1 \nbeta_{11}\,   \tilde k^2   +\frac23  \alpha_B^2 \tilde k^4 }{\mbeta_1+ \alpha_B^2\, \tilde k^2  } H^2  \sigma_{\rm m}   - \frac{\alpha_K}{\DD} \delta p_{\rm m}- 2 H \dot \sigma_{\rm m} \Bigg]\\
&\alpha_B^2 \tilde k^2 \left[ \Phi - \Psi \left( 1 + \alpha_T + \frac{ 2 \gamma_9  }{ \DD \alpha_B} \right) + \frac{\sigma_{\rm m}}{M^2} \right] 
 + \mbeta_1  \left[ \Phi - \Psi ( 1 + \alpha_T ) \frac{\gamma_1}{ \mbeta_1}  + \frac{\sigma_{\rm m}}{M^2} \right] = \nonumber \\
&\frac{\gamma_9}{ H^2  M^2}    \left[ \frac{\alpha_B}{\DD} \,  ( \delta \rho_{\rm m} -3 H q_{\rm m} )+  H M^2  \, \dot \Psi + H \frac{\alpha_K }{2 \DD} \, q_{\rm m}  -  H^2  \, \sigma_{\rm m}   \right] \;.
\end{align}
Here the parameters $\beta_a$, for which we have assumed again $\alpha_H=0$, are:\footnote{These parameters do not exactly correspond to those introduced in \cite{Bellini:2014fua}. Indeed, here the $\nbeta_a$ and $\Upsilon$ have been made dimensionless by dividing by the appropriate power of $H$ and, because of the different definition of $\alpha_B$, we have divided $\beta_1$ by a factor 4 so that $\mbeta_1{}^{(\rm here)} = \beta_1^{(\rm there)}/(4H^2)$. Moreover, we have corrected minor typos in the definitions of $\Upsilon$ and $\nbeta_7$. 
We thank the authors of Ref.~\cite{Bellini:2014fua} for having checked and privately agreed on these corrections.}
\begin{align}
\mbeta_1  & \equiv  \gamma_1 - \gamma_9 = - \alpha_K \frac{\rho_{\rm m} + p_{\rm m}}{4 H^2 M^2 } - \frac12 \DD \left( \frac{\dot H}{H^2} + \alpha_T -  \alpha_M   \right)  \;, \label{beta1}\\
\nbeta_2 & \equiv 2(2+\alpha_M) + 3 \Upsilon  \;, \\
\nbeta_3 & \equiv 3+\alpha_M + \frac{\alpha_B^2}{H  \DD} \left(  \frac{ \alpha_K}{\alpha_B^2} \right)^{\hbox{$\cdot$}} \;,\\
\nbeta_4 & \equiv (1+\alpha_T) \big[ 2 \dot H /H^2 + 3 (1+ \Upsilon) + \alpha_M \big] + \dot \alpha_T /H \;,  \\
\nbeta_5 & \equiv c_s^2 - \frac{ 2 \alpha_B (\beta_3- \beta_2)}{\DD}
+\frac{\alpha_B^2 }{\beta_1} (1+\alpha_T) (\beta_3 - \beta_2 ) +\frac{\alpha_B^2 \beta_4}{\beta_1}
 \;, \\
\nbeta_6 & \equiv \nbeta_7 - 2 \frac{\alpha_B(\nbeta_3 -\nbeta_2)}{\DD} \;, \\
\nbeta_7 & \equiv c_s^2 + 2 \frac{\alpha_B^2 (1+ \alpha_T ) +  \alpha_B (\alpha_T - \alpha_M)}{ \DD} \;,  \\
\nbeta_8 & \equiv \nbeta_9 -\frac{(\alpha_K -6 \alpha_B )(\nbeta_3 -\nbeta_2)}{\DD} \;, \\
\nbeta_9 & \equiv   -(1+3 c_s^2 +  \alpha_T)  + \frac{\alpha_B^2}{H  \DD} \left(  \frac{ \alpha_K}{\alpha_B^2} \right)^{\hbox{$\cdot$}} \;, \\
\nbeta_{10} & \equiv  - 6 (1+ \Upsilon) - 4 \dot H/H^2 \;, \\
\nbeta_{11} & \equiv\frac{2}{3} - 2 \frac{\alpha_B^2}{\beta_1}\big[ (2 - \alpha_M) +2  \dot H/H^2  \big] - 2 \frac{\alpha_B^4}{ \beta_1 H \DD} \left(  \frac{ \alpha_K}{\alpha_B^2} \right)^{\hbox{$\cdot$}} \; ,
\end{align}
with
\be
\begin{split}
12 \mbeta_1 H^3 M^2 \Upsilon \equiv  & \ 2 {\DD} M^2  \Big\{ \big[ \dot H +  (\alpha_T - \alpha_M) H^2 \big]^{\hbox{$\cdot$}} +  (3+\alpha_M) H \big[ \dot H +  (\alpha_T - \alpha_M) H^2 \big]  \Big\} \\
& + \alpha_K  {\dot p_{\rm m}} - ( {\rho_{\rm m} + p_{\rm m}})  H (\alpha_K - 6 \alpha_B) (\alpha_T - \alpha_M) + 6 ( {\rho_{\rm m} + p_{\rm m}})  \frac{\alpha_B^4}{  \DD} \left(  \frac{ \alpha_K}{\alpha_B^2} \right)^{\hbox{$\cdot$}}    \;.
\end{split}
\ee

\end{document}